\documentclass[journal=jacsat,manuscript=article]{achemso}

\usepackage[version=3]{mhchem} 
\usepackage{tabulary}
\usepackage[T1]{fontenc}       
\usepackage{arydshln}
\usepackage{multirow}
\usepackage{MnSymbol}
\usepackage{amsfonts}
\usepackage{makecell}

\usepackage{color}

\newcommand{\mstitle}{A streamlined, general approach for computing ligand binding free energies and its application to GPCR-bound cholesterol}

\newcommand{\msauthors}{
\author{Reza Salari}
\affiliation[Rutgers University]{Center for Computational and Integrative Biology, Rutgers University, Camden, NJ, USA}
\alsoaffiliation[Rutgers University]{Department of Physics, Rutgers University, Camden, NJ, USA}
\author{Thomas Joseph}
\affiliation[University of Pennsylvania Perelman School of Medicine]{Department of Anesthesia and Critical Care, University of Pennsylvania Perelman School of Medicine, Philadelphia, PA, USA}
\alsoaffiliation[Rutgers University]{Center for Computational and Integrative Biology, Rutgers University, Camden, NJ, USA}
\author{Ruchi Lohia}
\affiliation[Rutgers University]{Center for Computational and Integrative Biology, Rutgers University, Camden, NJ, USA}
\author{J\'{e}r\^{o}me H\'{e}nin}
\affiliation[CNRS]{Laboratoire de Biochimie Th\'{e}orique, Institut de Biologie Physico-Chimique, CNRS, Paris, France}
\alsoaffiliation[Joint]{These authors contributed equally to supervising this work.}
\author{Grace Brannigan}
\email{grace.brannigan@rutgers.edu}
\affiliation[Rutgers University]{Center for Computational and Integrative Biology, Rutgers University, Camden, NJ, USA}
\alsoaffiliation[Rutgers University]{Department of Physics, Rutgers University, Camden, NJ, USA}
\alsoaffiliation[Joint]{These authors contributed equally to supervising this work.}
}

\newcommand{\jerome}{}
\newcommand{\grace}{}

\newcommand{\ligandtype}{L}
\newcommand{\solventtype}{S}
\newcommand{\proteintype}{R}
\newcommand{\ligandproteintype}{RL}

\newcommand{\focc}{p_{\mathrm{occ}}}
\newcommand{\funocc}{p_{\mathrm{unocc}}}
\newcommand{\fbound}{p_{\mathrm{bound}}}
\newcommand{\ffree}{p_{\mathrm{free}}}

\newcommand{\ntotL}{N_{\ligandtype}}

\newcommand{\ntotLUnit}{n_{\ligandtype}}

\newcommand{\nfreeLUnit}{\ntotLUnit\ffree}

\newcommand{\ntotP}{N_{\proteintype}}

\newcommand{\ntotSUnit}{n_{\solventtype}}

\newcommand{\nsolvent}{n_{\solventtype}}

\newcommand{\cholfrac}{x_{\mathrm{CHOL}}}
\newcommand{\shortcholfrac}{x}
\newcommand{\conc}[1]{[#1]}
\newcommand{\concL}{\mathcal{\ligandtype}}
\newcommand{\concV}{\mathcal{V}}
\newcommand{\unitV}{{v}}
\newcommand{\totV}{{V}}

\newcommand{\concTotligand}{\conc{\ligandtype}_\mathrm{tot}}
\newcommand{\concTotprotein}{\conc{\proteintype}_\mathrm{tot}}
\newcommand{\concTotligandArea}{\conc{\ligandtype}_\mathrm{tot}^\mathrm{area}}
\newcommand{\concTotproteinArea}{\conc{\proteintype}_\mathrm{tot}^\mathrm{area}}
\newcommand{\act}[1]{\{#1\}}

\newcommand{\bphase}{b}
\newcommand{\gphase}{g}
\newcommand{\pphase}{r}

\newcommand{\grestraint}{\circ}
\newcommand{\occrestraint}{{\filledsquare}}
\newcommand{\norestraint}{{\smallsquare}}
\newcommand{\xrestraint}{}
\newcommand{\brestraint}{=}
\newcommand{\prestraint}{\smalltriangleup}
\newcommand{\Um}{U^{\brestraint}}
\newcommand{\Uz}{U_{z}}
\newcommand{\Utheta}{U_{\theta}}

\newcommand{\state}[2]{  {#1}^#2}

\newcommand{\pocc}{\focc}
\newcommand{\punocc}{\funocc}
\newcommand{\probrat}[1]{\kappa_{#1}}

\newcommand{\solvrat}[2]{P{}_{#2}}
\newcommand{\solvratD}[2]{\left. \partial_{\concL}{P_{\concL}}\right |_{#1}}

\newcommand{\solvratLnD}[2]{\left. \partial_{\concL}{\ln {P_{\concL}}}\right |_{#1}}

\newcommand{\zprobrat}{\frac{Z_\pphase^\occrestraint} {Z_\bphase^{\norestraint}} }
\newcommand{\zratOne}{\frac{Z_\bphase^\brestraint} {Z_\bphase^{\norestraint}}}
\newcommand{\zratTwo}{\frac{Z_\bphase^- Z_\gphase^\brestraint} {Z_\bphase^\brestraint}}
\newcommand{\zratThree}{\frac{Z_\gphase^\grestraint} {Z_\gphase^\brestraint}}
\newcommand{\zratFour}{\frac{Z_\gphase^{\grestraint\prestraint}} {Z_\gphase^\grestraint}}
\newcommand{\zratFive}{\frac{Z_\pphase^\prestraint} {Z_\gphase^\prestraint Z_\bphase^{-\xrestraint}}}
\newcommand{\zratFiveB}{\frac{Z_\pphase^\prestraint} {Z_\bphase^{-}}}

\newcommand{\zratOneP}{\left(\zratOne\right)}
\newcommand{\zratTwoP}{\left(\zratTwo\right)}
\newcommand{\zratThreeP}{\left(\zratThree\right)}
\newcommand{\zratFourP}{\left(\zratFour\right)}
\newcommand{\zratFiveP}{\left(\zratFive\right)}

\newcommand{\vx}{\mathbf{x}}
\newcommand{\vX}{\mathbf{X}}

\msauthors

\title{\mstitle}

\abbreviations{IR,NMR,UV}

\begin{document}

\begin{tocentry}

\includegraphics[width=9cm]{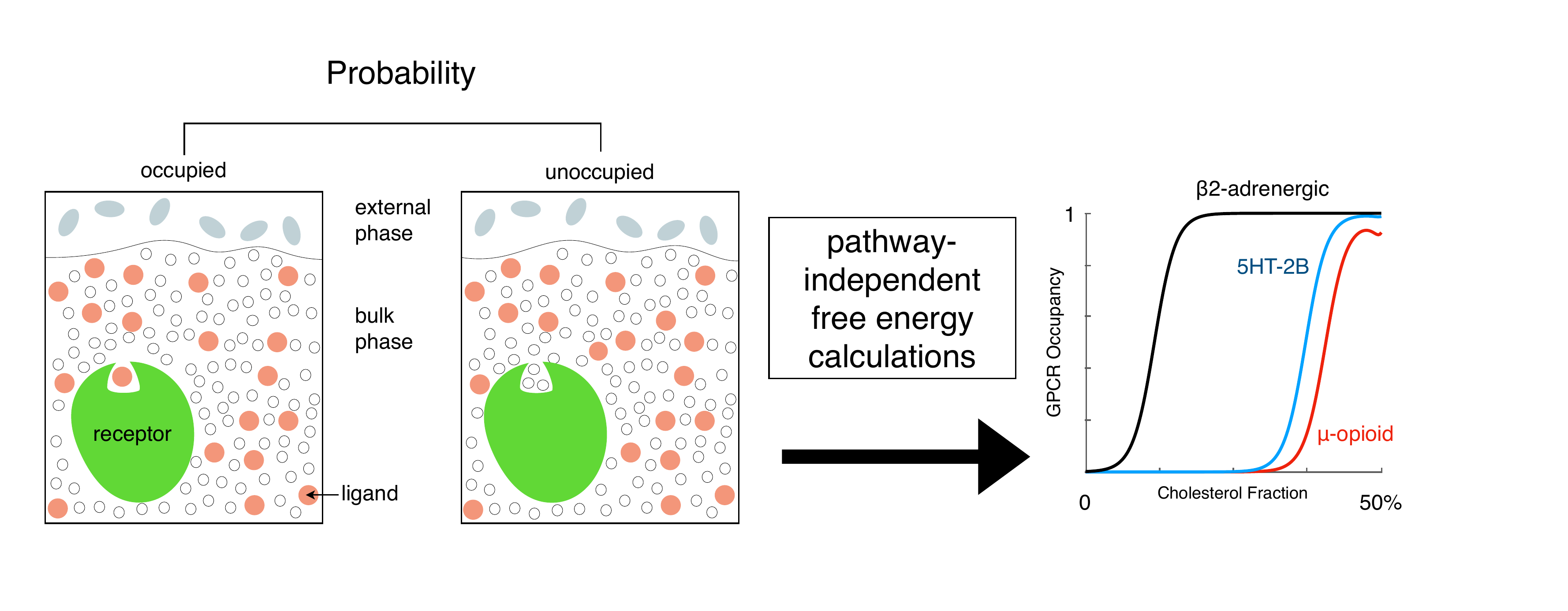}

\end{tocentry}

\begin{abstract}
The theory of receptor-ligand binding equilibria has long  been well-established in biochemistry, and was primarily constructed to describe dilute aqueous solutions.
Accordingly, few computational approaches have been developed for making quantitative predictions of binding probabilities in environments other than dilute isotropic solution. Existing techniques, ranging from simple automated docking procedures to sophisticated thermodynamics-based methods, have been developed with soluble proteins in mind.
Biologically and pharmacologically relevant protein-ligand interactions often occur in complex environments, including lamellar phases like membranes and crowded, non-dilute solutions.
Here we revisit the theoretical bases of ligand binding equilibria, avoiding overly specific assumptions that are nearly always made when describing receptor-ligand binding.
Building on this formalism, we extend the asymptotically exact Alchemical Free Energy Perturbation technique to quantifying occupancies of sites on proteins in a complex bulk, including phase-separated, anisotropic, or non-dilute solutions, using a thermodynamically consistent and easily generalized approach that resolves several ambiguities of current frameworks. To incorporate the complex bulk without overcomplicating the overall thermodynamic cycle, we simplify the common approach for ligand restraints by using a single distance-from-bound-configuration (DBC) ligand restraint during AFEP decoupling from protein. DBC restraints should be generalizable to binding modes of most small molecules, even those with strong orientational dependence.
We apply this approach to compute the likelihood that membrane cholesterol binds to known crystallographic sites on 3 GPCRs (beta2-adrenergic, 5HT-2B, and mu-opioid) at a range of concentrations. Non-ideality of cholesterol in a binary cholesterol:phosphatidylcholine (POPC) bilayer is characterized and consistently incorporated into the interpretation. \grace{We find that the three sites exhibit very different affinities for cholesterol: the site on the adrenergic receptor is predicted to be high affinity, with 50\% occupancy for 1:$10^{9}$ CHOL:POPC mixtures.   The site on the 5HT-2B and mu-opioid receptor are predicted to be lower affinity, with 50\% occupancy for $1:10^{3}$ CHOL:POPC and $1:10^{2}$ CHOL:POPC respectively.  }
These results could not have been predicted from the crystal structures alone. 
\end{abstract}
\section{Introduction}

Over the past two decades numerous advancements have improved the accuracy and
precision of methods for calculating binding free energies.
Force field parameters for
ligands are now developed consistent with the parent force field using largely
automated tools.\cite{Malde2011, Vanommeslaeghe2012, Vanommeslaeghe2012a, Mayne2013, Lundborg2015, Zheng2016, Dodda2017}
Many studies have reported successful use of free energy calculation methods to reproduce or predict experimental binding affinities\cite{Woo2005a, Gilson2007, Chipot2008, Pohorille2010, Henin2010b, Gumbart2013}.
Typical applications involve predicting dissociation constants for ligands binding in dilute-isotropic solution. 
One of the most common such approaches is double decoupling using Alchemical Free Energy Perturbation (AFEP).   Although the method is theoretically exact and pathway independent, it is technically challenging to achieve convergence.  Several groups have also laid the groundwork for employing restraints to tremendously improve the convergence of free energsy calculations, \cite{Gilson1997a,Boresch2003,Wang2006} but even in the simple case of a ligand binding to  a protein in aqueous solution, designing, implementing, and correcting for such restraints is a delicate process.  

\grace{The theory associated with current approaches for calculating binding affinities targets an isotropic dilute solution, treated as an ideal mixture quantified on a volume concentration scale. 
Many affinity prediction problems originate from pharmacology and involve high-affinity binding of dilute ligands, but the increasingly used fragment-based drug design approach implies high concentrations of small, relatively low-affinity ligands.\cite{Chen2015}
In these contexts, non-ideality becomes a potential source of error, which is not currently addressed in the biomedical and pharmacological literature.
Binding from even an isotropic non-ideal bulk has not, to our knowledge, been previously addressed using molecular simulation methods for measuring affinities and predicting titration. }

\grace{Furthermore, many non-ideal biologically relevant situations involve a bulk solution that is a complex or phase-separated fluid, with a high effective ligand concentration (within the preferred phase) even if the overall ligand concentration is low. This scenario is not well-suited to the traditional underlying formalism for treatments of ligand binding.  
As an example, calculating the binding free energy of a lipid for a site on a membrane protein 
presents several additional challenges compared to solution calculations. Physiological membranes are frequently non-ideal mixtures composed of multiple species with significant abundances, and consequently cannot be treated as a dilute solution.  
Even if a dilute solution approximation is made, however, interpretation of calculated binding affinities is also complicated due to ambiguities in the concentration scale and appropriate standard state, since the typical solution standard state (1~M or 1 molecule per 1660~\AA$^3$) is not directly transferable to quasi-two-dimensional lipid bilayer systems.
(Some authors \cite{Zhang2006} have approximated the volume of the membrane by the volume of the hydrophobic region of the membrane, thus permitting the use of the volume-based solution standard state definition, but this approach is not wholly satisfying since it precludes straightforward comparison between affinity measurements in membranes of different thicknesses.)
Quantifying experimental conditions on the water concentration scale would require achieving measurable concentrations of ligand in the water phase, which may not be possible with the most hydrophobic ligands. }

\grace{In a computational context, free energy calculations with respect to a non-isotropic bulk often introduce additional technical challenges already encountered for large or flexible ligands. Frequently ligand molecules in non-isotropic bulk may have a high aspect ratio and a strong orientation dependence in binding: simple center-of-mass restraints are too simplistic to sufficiently enhance convergence because they leave orientational and conformational degrees of freedom to be extensively sampled during the alchemical simulations. Conversely, more complete restraining schemes require a number of simulations to calculate the free energies of adding successive layers of restraints to the necessary degrees of freedom.\cite{Gumbart2013} In the example of a lamellar membrane environment, the distribution of a given lipophilic ligand is non-uniform across the membrane thickness, and superficial binding sites for lipids are not easily described using current restraint schemes. }

In this study, we present a general methodological framework for using AFEP calculations to calculate binding free energies for ligands  in a complex bulk, including quasi-2D lamellar systems as well as 3D solutions, and accounting for non-ideal and non-dilute mixtures. To make this tractable, we propose several adjustments to current practice: 
\begin{enumerate}
\item Extending the formalism for interpreting non-dilute and/or non-3D mixtures for comparison of simulation and empirical results, emphasizing physical observables rather than variables of arbitrary dimension.
\item Estimation of a concentration-response curve including non-ideality of the bulk by multiple bulk decoupling calculations at a range of concentrations.
\item Substantial simplification of the restraint scheme for a bound ligand.
\end{enumerate}
As an example application, we calculate the probability of binding of cholesterol from a POPC bilayer with 0-50\% cholesterol
to crystallographic sites on three different G-protein coupled receptors (GPCRs) : $\beta_{2}$-adrenergic, $\mu$-opioid, and 5-HT2 (serotonin) receptors. Function and organization of GPCRs is highly sensitive to cholesterol{\cite{Hanson2008, Oates2011, Venkatakrishnan2013}, multiple crystal structures show cholesterol in various binding modes for several classes of GPCR (reviewed recently in ref.~\citenum{Gimpl2016})   and interactions of cholesterol with GPCRs have received considerable interest from computational studies.\cite{Grossfield2011, Lee2012, Sengupta2012, Prasanna2014, Sengupta2015} To our knowledge, all computational estimates of the occupation probability have been determined  based on residence-times from equilibrium molecular dynamics simulations (either atomistic or coarse-grained), which are limited by residence times and lipid diffusion times that are similar to or longer than accessible simulation timescales.
In that approach, estimating concentration dependence requires separate simulations of the protein for every concentration.
Using our approach, we present probabilities of occupation for a range of cholesterol concentrations (trace  to 50\%).

Our method also quantifies and explicitly incorporates non-ideality of cholesterol:POPC membranes; we find that they are well-described as quadratic binary mixtures over the range of compositions with stable lamellar phases, with an unfavorable enthalpy of mixing.  

\grace{This manuscript presents an underlying theory followed by a specific application. {\it Theory} is structured as follows: A generalized macroscopic treatment presented in {\it Macroscopic Framework} is followed by a consistent {\it Microscopic Framework} for use in the interpretation of simulations. The microscopic treatment is then decomposed into intermediate states that are practical for a double decoupling approach, but that also clarify interpretation of concentrations in non-isotropic bulk phases, in {\it Decomposition into Intermediate States}. Next we specify how to interpret results of the method in {\it  Predicting Ligand Titration}, inspired by a double decoupling framework and including a regular solution theory for non-ideality. Finally, in {\it Generic flat-well restraints}, we present an approach for simulations with minimal geometric ligand restraints that satisfy assumptions previously made in the formalism. }

\grace{An application of the method to predict occupancy of cholesterol binding to three different GPCRs is also presented. Details of the implementation for this specific system are provided in {\it Methods}.
{\it Results} demonstrates non-ideality of cholesterol in phospholipid bilayers, which is considered with a fit to regular solution theory, independent of binding to any protein.  The final outcome is a prediction of cholesterol titration curves for the three GPCRs.  } 
\section{Theory}
{\small
\begin{table}[!htp]
\caption{\label{tab:param} Symbols and notation repeated within the theory. }
\begin{center}
\begin{tabular}{|c|l|}
\hline
& Molecular Species \\
\hline
$\solventtype$ & solvent\\
$\ligandtype$ & ligand\\
$\proteintype$ & receptor (macromolecule)\\
$\ligandproteintype$ & receptor-ligand complex\\
\hline
\hline
& Absolute System Composition \\
\hline
$\ntotL$ & total number of ligand molecules\\
$\ntotP$ & total number of protein molecules\\
$\ntotLUnit$ & total number of ligand molecules (unitary system); $\ntotLUnit = \ntotL/\ntotP$\\
$\ntotSUnit$ & total number of solvent molecules (unitary system)\\
$\unitV$& volume of the unitary system\\
$\concV$ &  generalized reference ``volume''; usually $\unitV$ for isotropic solutions \\
$\concL$ &  bulk ligand generalized ``concentration'' ; $\concL = \ntotLUnit/\concV$\\
$\concTotligand$& total ligand volume concentration; $\ntotLUnit/\unitV$\\
\hline
\hline
& Equilibrium System Composition   \\
\hline
$\focc$ & fraction of binding sites that are occupied ($\equiv 1 - \funocc$) \\
$\fbound$ & fraction of ligand that is bound ($\equiv 1 - \ffree$) \\
$\conc{\ligandtype}$& free ligand volume concentration; $\nfreeLUnit/\unitV$\\
$\act{\ligandtype}$ & ligand activity\\

\hline
\hline
& Ligand environment \\
\hline
$\bphase$& ligand coupled to bulk phase\\
$\gphase$ & ligand in gas phase (decoupled) \\
$\pphase$& ligand coupled to protein \\
\hline
\hline
& Tagged ligand state and restraints \\
\hline
$\occrestraint$ & unrestrained, in binding site \\
$\norestraint$ & unrestrained, all ligands outside binding site \\
$-$ & tagged ligand removed (number of ligand molecules reduced by 1) \\
$\brestraint$ & under anisotropic restraint consistent with bulk phase \\
$\grestraint$ & under isotropic restraint enclosing volume $V^{\grestraint}$ \\
$\prestraint$ & under DBC (distance-from-bound-configuration) restraint\\ 
$\alpha$ & fraction of bulk phase enclosed by $\brestraint$ restraints\\
\hline
\hline
&Probabilities and Statistics \\
\hline
$Z_{i}^{j}$ & configurational partition function for a system with ligand coupled to phase $i$\\ 
& under restraints $j$ with composition further specified in Table \ref{tab:species}. \\
$\probrat{\concL}$ & generalized binding constant: $K_{A}\gamma_{\concL}\ffree$ \\
$\solvrat{o}{\concL}$ & bulk/gas partition coefficient of ligand at bulk concentration $\concL$\\
$h^{0}$ & mean-field enthalpy of mixing, vanishes in ideal mixture \\
$\solvrat{o}{}^{0}$ & bulk/gas partition coefficient in ideal mixture\\
\hline
\end{tabular}
\end{center}
\end{table}
}

There are two equivalent ways to describe the state in which a ligand non-covalently binds to a receptor:  1) a macroscopic framework, in which a new chemical species (the receptor-ligand complex) is created and two separate entities (the free ligand and the receptor) are annihilated or 2) a microscopic framework, in which the bound state refers to a certain spatial localization of the ligand molecule (within the receptor binding site).
While the former makes use of the existing familiar framework established for reaction equilibrium, the latter is more directly relatable to the calculations used in explicit computational methods.
Although these two conventions are equivalent, consistency within a given treatment is essential. We begin within the macroscopic framework for empirical relevance, then switch to the microscopic framework to rigorously relate experimental observables to quantities accessible from simulations.

\subsection{Macroscopic Framework}
In this framework commonly used by experimental scientists, for consistency with the usual treatment of covalent reactions, the receptor-ligand complex is treated as a new chemical species, whose formation requires the annihilation of a free ligand and an unliganded macromolecule:
\begin{equation}
\ligandtype + \proteintype \rightarrow \ligandproteintype  
\end{equation}

For a closed system with $\ntotL$ total ligand molecules, $\ntotP$ total receptor molecules, and a one to one binding stoichiometry, the system composition is specified by the concentration of each supramolecular species, $\conc{\ligandtype}$, $\conc{\proteintype}$, and $\conc{\ligandproteintype}$, at equilibrium, and the probabilities of a ligand being bound or free or a protein site being occupied or unoccupied are $\fbound$, $\ffree$, $\focc$ and $\funocc$ respectively.  The following definitions and identities also hold, where Table \ref{tab:param} elaborates on the meaning of each symbol.  
\begin{eqnarray}
   \concTotligand\equiv \frac{\ntotL}{V} &;&
     \concTotprotein\equiv \frac{\ntotP}{V}\\
\conc{\ligandtype} \equiv \concTotligand-\conc{\ligandproteintype} =\ffree\concTotligand&;& \conc{\proteintype} \equiv \concTotprotein-\conc{\ligandproteintype} =\funocc\concTotprotein \label{eq:free}\\
\fbound= 1-\ffree =\frac{\conc{\ligandproteintype}}{\concTotligand} &;&\focc=1-\funocc = \frac{\conc{\ligandproteintype}}{\concTotprotein} \\
\fbound\concTotligand= &\conc{\ligandproteintype} & = \focc \concTotprotein 
\end{eqnarray}

In a binding assay, $\concTotprotein$ would typically be fixed, $\concTotligand$ would be varied, and $\focc$ would be measured:
\begin{equation}
\focc = \frac{1}{1 +  \frac{\funocc}{\focc}} = \frac{1}{1 +  \left(\frac{\conc{\ligandproteintype}}{\conc{\proteintype}}\right)^{-1}}
\end{equation}

Three different issues make the natural independent variable $\concTotligand$ only indirectly related to the typical theory that is used to interpret the results (and thus to computational methods).  
We consider these complicating factors in turn.

\subsubsection*{Complication 1: Free vs Total}

By definition of $K_{A}$, in the infinitely-dilute limit, 
\begin{eqnarray}
\frac{\conc{\ligandproteintype}}{\conc{\proteintype}} &=& K_{A} ~{\conc{\ligandtype}} = K_{A} ~ \ffree~\concTotligand 
\end{eqnarray}
where the latter equality uses Eq. \ref{eq:free}.  It is common to carry out experiments under conditions of high excess ligand, corresponding to $\ntotLUnit = {\concTotligand}/{ \concTotprotein}  >> 1$, and then assume that $\ffree \sim 1>> \fbound$.
This assumption, however, is needlessly restrictive, and our general formalism does not depend on it.

\subsubsection*{Complication 2: Non-ideality; Concentration vs Activity} 

In the general case, regardless of dilution, the activities of the three species obey:
\begin{equation}
\frac{\act{\ligandproteintype}}{\act{\proteintype}} = K_{A} {\act{\ligandtype}}
\end{equation}

Here we assume either no interactions between receptors; or that interactions between receptors are not significantly affected by ligand binding, so that non-ideal contributions of occupied and unoccupied receptors cancel out:
\begin{eqnarray}
\frac{\conc{\ligandproteintype}}{\conc{\proteintype}} &\sim&
\frac{\act{\ligandproteintype}}{\act{\proteintype}} = K_{A} ~{\act{\ligandtype}} =  K_{A} ~\gamma_{\ligandtype}{\conc{\ligandtype}}\\
\frac{\conc{\ligandproteintype}}{\conc{\proteintype}} &=&K_{A} \gamma_{\ligandtype} \ffree\concTotligand \label{eq:linear}
\end{eqnarray}
In the laboratory, plotting $ \frac{\conc{\ligandproteintype}}{\conc{\proteintype}}$ vs $\concTotligand $ would provide an indication of whether the quantity $ {K_{A}}{} \gamma_{\ligandtype}\ffree$ was constant, as is frequently assumed.  It would not indicate, however, whether non-linearity originated from non-abundant ligand (Complication 1) or  non-ideality (Complication 2), but it can be expected that the former case would dominate at low concentrations and the latter case would play a role at higher concentrations.

We may write more concisely the dependence of receptor occupancy on total ligand concentration:
\begin{eqnarray}
\frac{\conc{\ligandproteintype}}{\conc{\proteintype}} = \frac{\focc}{\funocc}&=& \probrat{{\ligandtype}}\;\concTotligand \label{eq:ratio}
\end{eqnarray}
where we define an equilibrium coefficient  $\probrat{{\ligandtype}}$ with dimensions of inverse concentration :
\begin{eqnarray}
    \probrat{{\ligandtype}}&\equiv &  K_{A}  \; \gamma_{\ligandtype}\; \ffree
\end{eqnarray}
and that contains all the information related to interactions of the ligand with the protein or in solution.

We are motivated to define this new quantity because our formulation of AFEP yields a value for  $\probrat{{\ligandtype}}$ that can be used to predict experimental observables ($\focc$) for a given $\concTotligand$.
If the relevant experimental conditions are non-ideal with respect to the ligand, affinity predictions can be carried out for those conditions without estimating $K_{A}$ and $\gamma_{\ligandtype}$, which would require additional computational work. Thus we avoid the conventional and often artificial reference to infinitely dilute conditions whenever such conditions are not relevant.

\subsubsection*{Complication 3: Concentration dimensions} 

Once we release ourselves from a commitment to calculating $K_{A}$, we are also able to generalize the notion of the relevant concentration.
In membranes, for instance, either an area concentration or a mole fraction may provide the most direct relationship with empirical data and physiological mechanisms. In a well-defined but irregular phase (such as a solution of micelles), {$\ntotL$} and {$\ntotP$} may be the only accessible variables.

Assuming no binding cooperativity among receptors as stated above, a closed system containing $\ntotP$ receptor molecules behaves as $\ntotP$ non-interacting copies of a ``unitary'' system containing one receptor and possessing the same intensive properties, including the mole fractions of solvent and ligand molecules.
Ligand abundance may be measured as the number of ligand molecules in such a unitary system: $\ntotLUnit = \ntotL/\ntotP $.

From this point forward, our formalism uses 
a ``generalized total ligand concentration'' $\concL$
\begin{eqnarray}
  \concL&\equiv&  \frac{\ntotL}{\ntotP}\frac{1}{\concV}  = \frac{\concTotligand}{\concTotprotein}\frac{1}{\concV} = \frac{\ntotLUnit}{\concV} \\
  &=&\begin{cases} 
  {\concTotligand}~&\mbox{if~} \concV\equiv\frac{1}{\concTotprotein}\\
  {\concTotligandArea}~&\mbox{if~} \concV\equiv~\frac{1}{\concTotproteinArea}\\
  \ntotLUnit~&\mbox{if~} \concV\equiv1\\
  \end{cases}
  \end{eqnarray}
 where $\concV$ is a generalized volume with dimensions and magnitude for the specific application (note that because it is a unitary volume, it is actually intensive). A convenient choice is $\concV = \unitV ={1}/{\concTotprotein} $ (the volume of the unitary system),  because then $\concL = \frac{\ntotLUnit}{\unitV} =  \frac{\ntotL}{\totV}$ is the volume concentration of the ligand within the phase.  In other geometries,  it may make sense to set $\concV$ equal to the area of a unitary system (suitable for quasi-2D systems like membranes):
  or even just set $\concV = 1$ so that $\concL = \ntotLUnit$, the number of ligand molecules in the unitary system.  The equilibrium coefficient is also then generalized to
    \begin{eqnarray}
     \probrat{\concL}&\equiv &  \frac{\focc}{\funocc} \frac{1}{\concL} =  \frac{\focc}{\funocc} \frac{\concV}{\ntotLUnit} 
  \end{eqnarray}

Then explicitly:
\begin{eqnarray}
\focc & =&\frac{1}{1 + \punocc/\pocc}\\
&=& \frac{1}{1 + \frac{1}{\probrat{\concL}{\concL}}}
\label{eq:foccKappa}
\end{eqnarray}

       \begin{figure}[!ht]
      \centering
      \includegraphics[scale=0.5]{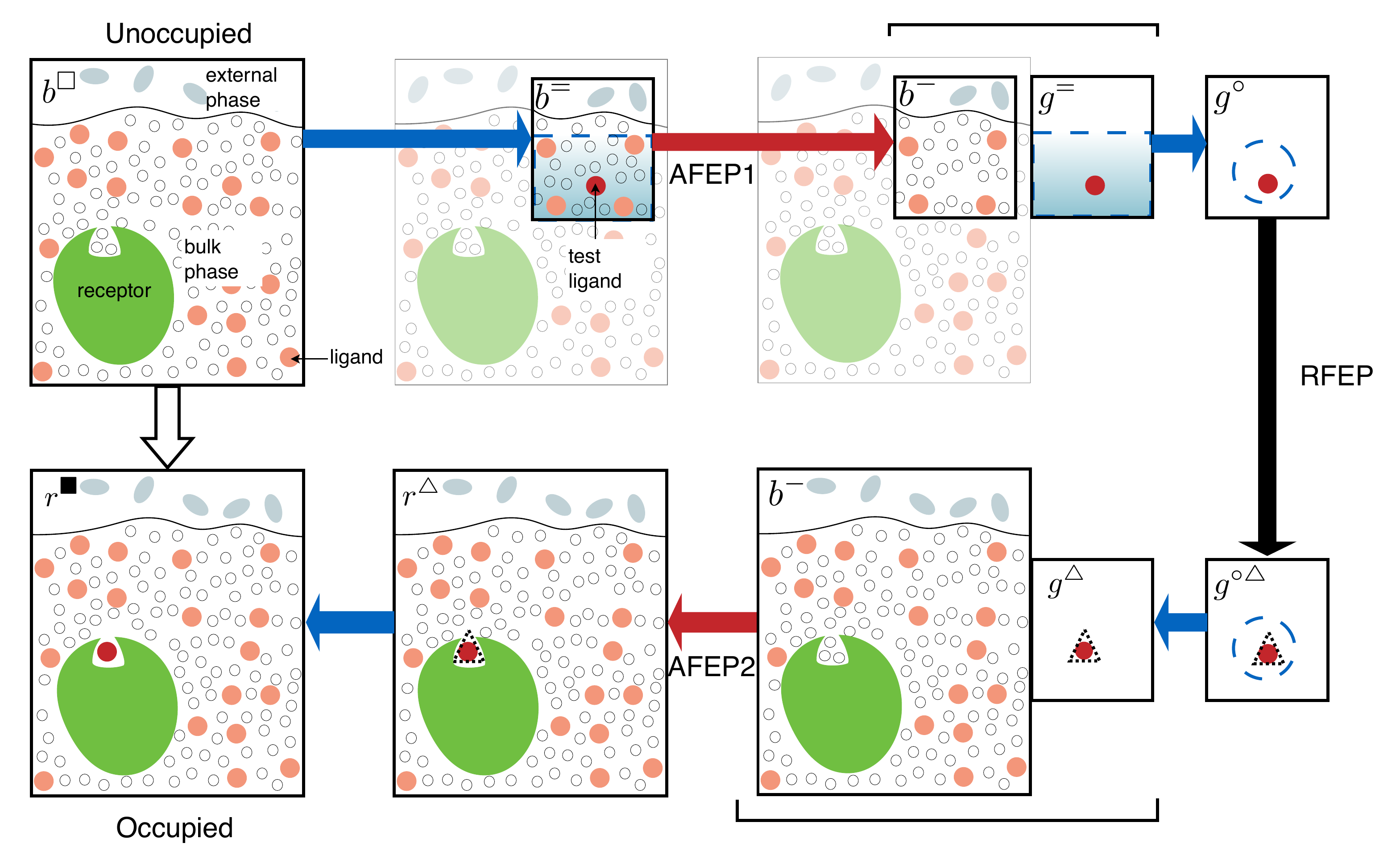}

\caption{{\bf Schematic presentation of the various intermediate states connecting the unoccupied and occupied states of the unitary system.} The receptor is green, molecules in the external phase are gray, bulk solvent and ligand molecules are white and pale red circles respectively.  The test ligand for decoupling from bulk (AFEP1) and from protein (-AFEP2) is shown as a bright red circle.  Blue dashed lines indicate the test ligand is subject to restraints with simple translational geometry, with the shaded gradient indicating possible orientational restraints for highly anisotropic bulk.  Black dashed lines indicate the test ligand is under DBC restraints.  Systems that are actually simulated are outlined in thicker line, whereas greyed-out components are included in the theory, but omitted from numerical simulations when doing so does not alter the results. Color of arrows represents technique: alchemical free energy perturbation (AFEP, red), restraint free energy perturbation (RFEP, black), analytical (blue). Braces connect pairs of decoupled systems that are sampled simultaneously in alchemical simulations.
}
      \label{fig:cycle}
    \end{figure}
    
\subsection{Microscopic Framework}

The \emph{unitary system} connects the macroscopic framework to a microscopic framework that best describes molecular simulations.  It is defined as a bulk liquid phase containing a single receptor with a binding site and two types of small molecules, $\ligandtype$ and $\solventtype$, with composition 1:$\ntotLUnit$:$\nsolvent$ for receptor:$\ligandtype$:$\solventtype$, as illustrated in Figure \ref{fig:cycle}. 
This unitary system is equivalent to a unit cell in a ``lattice'' binding model. \cite{Phillips2012}
Although we refer to the solvent as species $\solventtype$ for the sake of simplicity, it may consist of an arbitrary mixture.
This liquid (or liquid crystal) phase may coexist with other phases, and we make no assumption about its geometry, so that it may represent membranes, micelles, or other aggregates.

In the microscopic perspective, chemical entities are stable, yet a supramolecular receptor-ligand complex may exist when the coordinates of both molecules satisfy certain geometric criteria.
Appropriate criteria vary across applications, making a robust and general implementation challenging thus far.
We postpone making specific remarks about these geometric criteria, but define them in the abstract as separating the statistical ensemble of our unitary system into two macrostates:  either one ligand satisfies the criteria, with probability $\pocc$, or no ligand does, with probability $\punocc$.

To determine the ratio $\pocc/\punocc$, we use an approach that  generalizes the double decoupling method of Gilson et al.\cite{Gilson1997a}, beginning with the overall ratio of partition functions between the state ${{\pphase}^{\occrestraint}}$, when one ligand satisfies the structural requirements for occupancy, and the state ${\bphase}^{\norestraint}$ in which that ligand is in the bulk environment and no other ligands occupy the binding site,
 \begin{equation}
\probrat{\concL}\concL=\frac{\pocc}{\punocc} = \frac{Z_{\pphase}^{\occrestraint}}{Z_{\bphase}^{\norestraint}}
 \label{eq:equivalent}
 \end{equation}
 where $Z^{\norestraint}_{\bphase}$ and $Z^{\occrestraint}_{\pphase}$ are the configurational partition functions for the $\state{\bphase}{\norestraint}$ and $\state{\pphase}{\occrestraint}$ states respectively. 

A summary of notation and parameter definitions can be found in Table~\ref{tab:param}, and a relevant thermodynamic cycle is depicted in Figure~\ref{fig:cycle}.
The AFEP double-decoupling scheme obtains an excess free energy of transfer from two main alchemical simulations, decoupling the ligand from two different locations within this larger system, the binding site and the bulk, and these two calculations can be reasonably carried out in smaller systems modeling the local environment only.

An inherent complication of the AFEP method, which involves gradually
decoupling a ligand from its environment over the course of an MD simulation,
is that a very weakly or non-coupled ligand will spend significant simulation
time exploring configurational states that are highly improbable in the coupled
state (for example, unbinding from the site and diffusing freely in the
simulation box). Extensive sampling of these states drastically reduces
efficiency of the calculation and quality of convergence, since these
improbable states make negligible contributions to the binding affinity.
Typically, restraints\cite{Boresch2003,Gilson1997a,Wang2006,Gumbart2013} are
applied to the ligand throughout the calculation to restrict sampled
configurations to those likely to be found in the coupled state, but the final
calculation of the binding affinity must include corrections for any
contributions of these restraints, complicating the thermodynamic cycle
represented by the overall calculation.  Our general approach involves simplification of the usual restraint scheme for bound ligands, and introduces more complex bulk phases which may have their own applied restraints.   Although the simplification was motivated by the need to streamline the overall process when including complex bulk phases, it is likely to make calculations in aqueous solution even more straightforward.  

\renewcommand{\arraystretch}{1.5}
\begin{table}[htp]
\begin{center}
\begin{tabular}{c|cccc|c|l}
State& \makecell{Tagged Ligand \\ Environment} &\makecell{Total Ligand \\Molecules }
&Restraints & \makecell{Restraint\\ Symbol} &Z& Technique\\
\hline\hline
$\state{\bphase}{\norestraint}$&bulk &$\ntotLUnit$&none$^{\dagger}$  &&$Z^{\norestraint} $&\\ 
\cline{1-7}
$\state{\bphase}{\brestraint}$ &bulk&$\ntotLUnit$& bulk$^{\dagger}$&$\brestraint$&$Z_{\bphase}^{\brestraint}$&\multirow{2}{*}{AFEP}
\\
\cline{1-6}
$\state{\bphase}{{\xrestraint-}}$&N/A  &$\ntotLUnit -1 $&  none$^{\dagger}$&&\multirow{2}{*}{$Z_{\bphase}^{-}Z_{\gphase}^{\brestraint}$} &\\
$\state{\gphase}{\brestraint}$ &gas-phase&	1 & bulk$^{\dagger}$&$ \brestraint$ & \\
				\cline{1-7}
$\state{\gphase}{\grestraint}$					   &gas-phase&1&	coarse &$\grestraint$&  $Z_{\gphase}^{\grestraint}$&\multirow{2}{*}{RFEP}\\
				\cline{1-6}
$\state{\gphase}{{\grestraint\prestraint}}$	  &gas-phase&	1& coarse  and DBC &$\grestraint\prestraint$&$Z_{\gphase}^{\grestraint\prestraint}$ \\
				\cline{1-7}
$\state{\gphase}{{\prestraint}}$ &gas-phase &1&  DBC & $\prestraint$&\multirow{2}{*}{$Z_{\gphase}^{\prestraint}$$ Z^{-\xrestraint}_{\bphase}$}  \\
$\state{\bphase}{{\xrestraint-}}$&N/A&$\ntotLUnit -1$& exclusion && &\multirow{2}{*}{AFEP}\\
\cline{1-6}
$\state{\pphase}{{\prestraint}}$&receptor&$\ntotLUnit$&DBC & $\prestraint$ &$Z_{\pphase}^{\prestraint}$\\
\cline{1-7}
$\state{\pphase}{\occrestraint}$ &receptor&$\ntotLUnit$  & none &  & $Z^{\occrestraint}$
\end{tabular} \end{center}
\caption{Properties of systems referenced in Figure~\ref{fig:cycle}.
This follows the microscopic framework in which only covalent bonds define a molecule. Separate systems are divided by solid lines, while isolated gas-phase systems associated with a unitary system are in an additional, un-separated row lines; column noted $Z$ contains notation for the configurational partition function for the two-system state.
Calculating $Z^\norestraint / Z^\occrestraint$ is the primary goal of the proposed implementation, and the 3 calculations that must be done computationally are noted in  the ``Technique'' column.  Other partition function ratios can be calculated analytically or cancel out.
\\ \\
$^{\dagger}$ Exclusion restraints are technically also required, but imposed automatically if simulation does not include protein.
}
\label{tab:species}
\end{table}

\subsection {Decomposition into Intermediate States}

The double decoupling method requires considering states in which a single ligand is interacting with 1) preferred bulk phase (${\bphase}$ subscript), 2) ideal gas phase (${\gphase}$ subscript), or 3) the environment associated with the protein binding site (${\pphase}$). For convergence purposes it is frequently advisable to impose restraints that confine the ligand to these environments.  We describe here a minimum set of such restraints, which confine the ligand to one section of the bulk (${\brestraint}$ restraints), ``coarse'' site restraints that confine it in the general region of the binding site (${\grestraint}$), and finer Distance-from-Bound-Configuration (DBC or ${\prestraint}$ ) restraints that restrict its RMSD from the bound configuration.  For the method to correspond to the binding affinity it is also essential that the site remain unoccupied by identical ligand molecules that may also be in solution, requiring explicit exclusion restraints at high ligand concentrations.   As described in Methods, these restraints can be designed so they do no significant work when the ligand is fully coupled to the associated phase.

Relevant combinations of these restraints and coupling schemes for the computational method to calculate the unitary binding probability ratio $\pocc/\punocc$ are shown in Figure~\ref{fig:cycle}, and more details on the composition of various states are given in Table \ref{tab:species}.  Although these states can be connected in a consistent thermodynamic cycle, we do not explicitly do so to derive the method, due to several terms that cancel.   

Instead, to obtain a form for $\frac{Z_{\pphase}^{\occrestraint}}{Z_{\bphase}^{\norestraint}}$ practical for calculation, we multiply and divide $\frac{Z_{\pphase}^{\occrestraint}}{Z_{\bphase}^{\norestraint}}$ by the partition function associated with each state in Figure~\ref{fig:cycle}:
\begin{equation}
\label{eq:zratio_all_terms}
\frac{Z_\pphase^\occrestraint} {Z_\bphase^\norestraint} =
\left(\frac{Z_\bphase^\brestraint} {Z_\bphase^\norestraint}\right)
\left(\frac{Z_\bphase^- Z_\gphase^\brestraint} {Z_\bphase^\brestraint}\right)
\left(\frac{Z_\gphase^\grestraint } { Z_{\bphase}^{-} Z_\gphase^\brestraint}\right)
\left(\frac{Z_\gphase^{\grestraint\prestraint}} {Z_\gphase^\grestraint}\right)
\left(\frac{Z_\gphase^\prestraint Z_{\bphase}^{-\xrestraint}} {Z_\gphase^{\grestraint\prestraint}}\right)
\left(\frac{Z_\pphase^\prestraint} {Z_\gphase^\prestraint Z_\bphase^{-\xrestraint}}\right)
\left(\frac{Z_\pphase^\occrestraint} {Z_\pphase^\prestraint}\right)
\end{equation}
In the right hand side, the grouping of terms is inspired by transitions shown in Figure~\ref{fig:cycle}, retaining ratios that can be conveniently calculated and omitting some terms that cancel out.

Equation~\ref{eq:zratio_all_terms} may be simplified by making the following assumptions:
${Z_{\pphase}^{\grestraint\prestraint}} \sim {Z_{\pphase}^{\prestraint}}$ (coarse $\grestraint$ restraints are broader than the DBC $\prestraint$ restraints, and hence negligible when superimposed onto them),
and
${Z_{\pphase}^{\prestraint}}\sim { Z_{\pphase}^{\occrestraint}}$ (effects of DBC restraints in the bound, coupled state are negligible). Our approach for designing restraints that satisfy these assumptions are in {\it Simulation Methods}. 
This leads to:
\begin{equation}
\label{eq:zratio_simplified}
\probrat{\concL} \concL =\zprobrat =
\zratOneP\zratTwoP\zratThreeP\zratFourP\zratFiveP
\end{equation}

All terms in Eq.~\ref{eq:zratio_simplified} can be computed either numerically or analytically for a given concentration $\concL$, and $\probrat{\concL}\concL$ can then be substituted into Eq.~\ref{eq:foccKappa} to calculate the probability of site occupation at $\concL$.  

Usually, the first and third term can be estimated analytically, as shown in the next section, {\it Predicting Ligand Titration}.  The first term $\frac{Z_\bphase^\brestraint} {Z_\bphase^\norestraint}$ represents the cost of imposing possible bulk-phase restraints ($^\brestraint$); depending on implementation details, this may be unity or a simple ratio of volumes. The third term, $\frac{Z_{\gphase}^{\grestraint}}{Z_{\gphase}^{\brestraint}}$ yields the cost of switching from the $\brestraint$ restraint system to the $\grestraint$ system; if both are flat wells, it is a ratio of configuration space volumes; if in addition they are purely translational, it is again a ratio of 3-dimensional volumes.

The fourth term, $\frac{Z_{\gphase}^{\grestraint\prestraint}} {Z_{\gphase}^{\grestraint}}$ is a correction for the DBC restraints in the gas phase, represented by the RFEP arrow in Figure \ref{fig:cycle}.
\jerome{
As with other flat-well potentials, a stiff DBC restraint scales the partition function of the decoupled ligand by a volume ratio.
The 3n-dimensional volume enclosed by a DBC restraint is not regular or analytically calculable, but can be estimated numerically, using restraint free energy perturbation (RFEP) simulations coupled with a free energy estimator such as thermodynamic integration (TI) or Overlap Sampling methods~\cite{LuN2003}.
In practice, convergence is improved by releasing the DBC restraints in RFEP but maintaining superimposed regular (usually spherical) flat-well center of mass restraints ($\grestraint$ restraints) that enclose the DBC volume to calculate $\zratFour$, then correcting for those restraints analytically via $\zratThree$.}

The second $\zratTwoP$ and fifth terms $\zratFiveP$ correspond to the AFEP1 and AFEP2 decoupling steps in Figure \ref{fig:cycle}, respectively.

 \newcommand{\vratOne}{$\frac{V^{\grestraint}}{V^{\brestraint}}$}
\newcommand{\nratOne}{$\frac{\ntotLUnit V^{\brestraint}}{V^{\norestraint}}$}
\newcommand{\nratTwo}{$\ntotLUnit \alpha$}
\newcommand{\omegaratOne}{$ \frac{\Omega^{\grestraint}}{\Omega^{\brestraint}}$}
\newcommand{\omegaratTwoP}{$ \left(\frac{\Omega^{\grestraint}/V^{\grestraint}}{\Omega^{\brestraint}/V^{\brestraint}} \right)$}
\newcommand{\omegaratTwo}{$ \frac{\Omega^{\grestraint}/V^{\grestraint}}{\Omega^{\brestraint}/V^{\brestraint}}$}
\newcommand{\solvratOne}{$\left.\solvrat{o}\right|_0$}
\newcommand{\solvratTwo}{\left.\solvrat{o}\right|_{\concL_{0}}\times\left(1+\solvratLnD{\concL_0}{o}\left({\concL} - {\concL_0}\right)\right)}
\newcommand{\solvratTwoB}{$\left.\solvrat{o}\right|_0~\left(1+\solvratLnD{0}{o}\concL\right)$}
\newcommand{\solvratThree}{$\left.\solvrat{o}\right|_{\concL_0}\times\left(1 + \concL_0 \solvratLnD{\concL_0}{o} \right)$}
\newcommand{\solvratThreeB}{$\left.\solvrat{o}\right|_0 $}
\newcommand{\solvratFour}{$\frac{\left.\solvrat{o}\right|_{\concL_0}+\solvratD{\concL_0}{o}\left({\concL} - {\concL_0}\right)}{\left.\solvrat{o}\right|_{\concL_0} + \concL_0 \solvratD{\concL_0}{o} }$} 
\newcommand{\solvratFiveNum}{1+{\solvratLnD{\concL_0}{o}}\left(\concL - {\concL_0}\right)}
\newcommand{\solvratFiveNumB}{1 - {\solvratLnD{\concL_0}{o}}\concL_0}
\newcommand{\solvratFiveDenom}{1 + {\solvratLnD{\concL_0}{o}}\concL_0}
\newcommand{\idealratOne}{$\zratFiveB\concL$ }

\subsection{Predicting ligand titration}

Experiments frequently involve titrating ligand, and for successful comparison it is desirable to predict the binding probability $\pocc$ for a range of ligand concentrations.
Such concentration effects (even among non-interacting receptors) have two distinct origins.
Increasing the number of ligand molecules $n_{L}$ in a unitary system increases the probability that any ligand will bind; this leads to the ``ideal gas'' concentration dependence.
For solutions with high ligand concentrations, ligand-ligand interactions are non-vanishing and contribute to the cost of solvation within the bulk; this is (to lowest order) a function of the typical number of interacting ligand-ligand pairs, as well as a function of the concentration and the geometry of the bulk.  
Thus, choosing the most relevant concentration scale for the ligand can be difficult, especially in phase-separated systems.

To our knowledge, most implementations of double decoupling to date have assumed a bulk that is a dilute, isotropic solution. It is possible to calculate $\pocc$ directly for any given bulk ligand concentration by simply decoupling from a bulk system at that concentration, but here we summarize a natural approach for incorporating ligand concentration effects into the formalism we have introduced.  

\renewcommand{\arraystretch}{2.0}
\begin{table}[htp]
\caption{Effects of bulk phase and geometry on concentration dependence of each term in Eq. \ref{eq:zratio_simplified}, for an ideal gas of ligand, an isotropic solution of ligand, and an anisotropic phase containing both ligand and receptor. Empty cells indicate the ideal gas value should be used. The factor of $\ntotLUnit$ in the first row originates from $\ntotLUnit$ indistinguishable ligands that can be chosen for restraining (and then decoupling).   The equilibrium coefficient $\probrat{\concL}$ is obtained by taking the product of the five ratios and dividing by $\concL = \ntotLUnit /\concV$. It is assumed that restraints are designed so $V^{\grestraint\prestraint} = V^{\prestraint}$ and that for the ideal gas and isotropic solution, the generalized reference volume is simply the unitary volume ($\concV = \unitV$). Although the final expression for $\probrat{\concL}$ for the anisotropic bulk appears considerably more complex, the additional terms can often be estimated analytically.}
\begin{center}
\begin{tabular}{c|c|ccc}
Process & Ratio & Ideal Gas & Isotropic Solution & Anisotropic Bulk \\
\hline
apply bulk restraints & $\zratOne $& $\frac{\ntotLUnit V^{\brestraint}}{\unitV}$ & & $\ntotLUnit \alpha$\\
decouple from bulk & $\zratTwo$& 1 &$ \frac{1}{\solvrat{o}{\concL}}$ & $\frac{1}{\solvrat{o}{\concL}}$ \\
switch to isotropic restraint & $\zratThree$& $\frac{V^{\grestraint}}{V^{\brestraint}}$ && $\frac{V^{\grestraint}}{V^{\brestraint}}$ $ \frac{\Omega^{\grestraint}}{\Omega^{\brestraint}}$\\
add DBC restraint & $\zratFour$&  $\frac{V^{\grestraint\prestraint}}{V^{\grestraint}} $\\
couple to receptor & $\zratFive$ & $\frac{Z_\pphase^\prestraint} {V^{\prestraint}  Z_\bphase^-}$\\
\hline
bind from bulk & $\probrat{\concL}$ &$\frac{ Z_\pphase^\prestraint}{Z_\bphase^-}$ & $\zratFiveB\frac{1}{\solvrat{o}{\concL}}$ & $\frac{Z_\pphase^\prestraint} { Z_\bphase^-}~ \frac{1}{\solvrat{o}{\concL}}~\frac{\Omega^{\grestraint}}{ \Omega^{\brestraint}}~\frac{\alpha \concV}{V^{\brestraint}}$
\end{tabular}
\end{center}
\label{tab:bulk}
\end{table}
Of the five ratios of partition functions in Eq. \ref{eq:zratio_simplified}, the first three are directly affected by the bulk ligand concentration and geometry, with distinct expected behavior depending on whether the ligand in bulk is  an ideal gas, or in solution that is dilute, non-dilute and/or anisotropic. The expectation for these expressions for each case is given in Table \ref{tab:bulk}. We consider the three deviations in turn.  

First, if multiple bulk phases are present, and the ligand is strongly localized in one of them (as a lipophilic ligand in a hydrated membrane),  $Z_\bphase^\norestraint$ is proportional to the volume of the accessible phase (states in which the ligand is outside its preferred phase will not contribute significantly to the partition function).  In practice, the volume of a phase with an irregular shape may not be easily characterized, but the bulk restraints (${\brestraint}$) can be defined to enclose a regular volume $V^{\brestraint}$ and estimating the fraction of the overall bulk phase enclosed in these restraints, as we do here, is more straightforward.
We may write:
\begin{equation}
\zratOne = \ntotLUnit \alpha \label{eq:bulkone}
\end{equation}
and estimate the volume ratio $\alpha$ using ligand numbers: 
\begin{equation}
\alpha \equiv \frac{V^{\brestraint}}{\unitV} = \frac{\ntotLUnit^{\brestraint} + \ntotSUnit^{\brestraint} }{\ntotLUnit + \ntotSUnit},
\end{equation}
where $\ntotLUnit^{\brestraint}$ and $\ntotSUnit^{\brestraint}$ are the number of ligand and solvent molecules enclosed by the restraint, respectively.
This amounts to estimating the unknown volume $\unitV$ of the bulk phase within the heterogeneous system containing the receptor based on the number density of the more symmetric ``pure'' bulk system (in the present application, a hydrated binary lipid bilayer).
The restraint volume $V^{\brestraint}$ is meaningful on the condition that the bulk restraints enclose a region of the bulk that maps to any other region under symmetry operations (such as a given fraction of a homogeneous phase, or one leaflet of a symmetric bilayer).

Second, we have introduced the bulk/gas partition coefficient $\solvrat{o}{\concL}$, which also captures any non-ideality of the bulk. The free energy of solvation for the ligand in a bulk solution with ligand concentration $\concL$ is $k_{B}T\ln \solvrat{o}{\concL}$, where 
\begin{eqnarray}
\frac{1}{\solvrat{o}{\concL}}& \equiv &   \zratTwo,\label{eq:P} 
\end{eqnarray} 
and the system is at standard temperature and pressure.  

In a binary mixture of two species A and B, with number fractions $x$ and $1-x$, the simplest deviation from ideal (known as a ``simple solution'', ``regular solution'',  or ``quadratic mixture'') yields the following chemical potential for species A: \cite{Rowlinson1982, Silver1985, Morris2008}
\begin{eqnarray}
\mu &= \mu^{0} + R T\ln{x}+ h^{0}(1-x)^{2}\label{eq:binary_chempot}
\end{eqnarray}
where $\mu^{0}$ is the chemical potential for species A in an infinitely dilute state, $h^{0} = u_{AB} - (u_{AA}+u_{BB})/2$ and $u_{AB}, u_{AA}, u_{BB}$ are the mean interaction energies for AB, AA, and BB pairs respectively. If the mean pair interaction energies do not vary with composition, $h^{0}$ will be a constant. 

Here, the natural generalized concentration is mole fraction, so we set $\concL = x$.  The bulk/gas partition coefficient can be determined by setting the chemical potential in the bulk $\mu_{\bphase}$ equal to the chemical potential in the gas phase  $\mu_{\gphase}$:
\begin{eqnarray}
 \mu_{\gphase}^{0} + RT\ln{x_{\gphase}} &=&  \mu_{\bphase}^{0} + RT\ln{x_{\bphase}}+ h^{0}(1-x_{\bphase})^{2}\\ 
\solvrat{o}{x} &= & \frac{x_{\bphase}}{x_{\gphase}} 
=  \solvrat{o}{0} e^{-h^{0}(1-x_{\bphase})^{2}/RT} \label{eq:binary}
\end{eqnarray}
where $x_{\bphase}$ and $x_{\gphase}$ are the mole fraction of Species A in the A:B mixture in liquid and gas phase, respectively, and $ \solvrat{o}{}^{0} \equiv e^{(\mu_{\gphase}^{0}  -   \mu_{\bphase}^{0})/RT}$ is the bulk/gas partition coefficient for species A in an ideal A:B mixture where $h^{0}=0$}. 
Simulations of binary mixtures of coarse-grained lipids have previously shown agreement over a full concentration range\cite{Brannigan2005b}. We find this model for $\solvrat{o}{x}$ agrees well with simulation results for membranes with less than 50\% cholesterol, and use it to predict cholesterol titration in the cholesterol/GPCR application presented below. Instability of the lamellar phase indicates the model must break down for cholesterol fractions greater than 50\%. 

Finally, for anisotropic phases in which ligand molecules have a strong orientational dependence (as for sterols in a bilayer membrane), bulk ($^{\brestraint}$) restraints may include an orientational component which is not neutral in the gas phase:
\begin{equation}
\frac{Z_{\gphase}^{\grestraint}}{Z_{\gphase}^{\brestraint}}  = \frac{V^{\grestraint}}{V^{\brestraint}}\frac{\Omega^{\grestraint}}{\Omega^{\brestraint}}\label{eq:bulkthree}
\end{equation}
where $\Omega^{\grestraint}$ and $\Omega^{\brestraint}$ include the phase-space volume for all non-translational degrees of freedom under the isotropic ($^{\grestraint}$) and bulk ($^{\brestraint}$) restraints, respectively.

\subsection{Generic flat-well restraints}
\label{sec:Implementation}

The implementation assumes restraint schemes that perform no work on the ligand coupled to either the bulk or the binding site. 
A common scheme for meeting this requirement is a ``flat-well'' potential. 
For ligand configurations that are likely in
the coupled state, the potential vanishes; for those that are highly unlikely, the potential increases steeply to approximate a hard wall
and return the ligand to the ``coupled'' configuration space. The
free energy cost for imposing such a flat-well restraint potential in the decoupled state is due only to the loss of entropy; for a flat-well potential in which the ligand is
restrained to a simple geometry, such as a sphere as in refs.~\citenum{Henin2010b,Woll2016b} or a
cylinder as in ref.~\citenum{LeBard2012}, this entropic contribution can be calculated
analytically. 

\jerome{
Here we use two types of flat-well restraints, depending on the coordinate $\xi$ they are applied to: either the center-of-mass distance of the ligand to the binding site, or DBC coordinates, described in detail in the next section.
Our flat-well restraints are half-harmonic wells:
}
\begin{equation}
U_\mathrm{FW}(\xi) =
\begin{cases}
0 &\text{if } \xi \leq \xi_\mathrm{max},\\
\frac{1}{2} \, k_{\xi}\,  (\xi - \xi_\mathrm{max})^2 &\text{if } \xi > \xi_\mathrm{max},
\end{cases} \label{eq:fw}
\end{equation}
which is parameterized by the threshold distance $\xi_{\mathrm{max}}$ and the force constant $k_{\xi}$.
$\xi_{\mathrm{max}}$ is chosen to make the restraint a flat well, that is, so that the equilibrium distribution of $\xi$ in the coupled state lies almost entirely below $\xi_\mathrm{max}$.
This ensures that imposing the restraint induces negligible bias on sampling in the coupled state.
$ k_{\xi}$ should be somewhat high to limit the space to be sampled in AFEP simulations; however, it must be low enough to preserve the stability and accuracy of the MD integrator.
In a Monte-Carlo simulation where this concern does not apply, a hard wall might be used instead.
Imposing this flat-well restraint on a decoupled ligand scales the partition function $Z$ by a volume ratio: $\frac{Z_{FW}}{Z} = \frac{V_{FW}}{V}$, where $Z_{FW}$ and $Z$ are the partition functions with and without the flat-well restraints, respectively, and $V_{FW}$ and $V$ are the configuration volumes accessible to the ligand in the presence and absence of restraints.  For a ligand coupled to protein, $\frac{Z_{FW}}{Z} =1$, while for a ligand coupled to bulk, $\frac{Z_{FW}}{Z} =\alpha$ where $\alpha$ is the fraction of the unitary system bulk enclosed by the restraint.

\subsection{Distance-to-bound-configuration coordinate}

\jerome{
In the past,\cite{Wang2006,Ge2010,Mobley2006} in
order to improve sampling and convergence of the bound ligand in the free energy
calculations up to 7 types of restraints (3 translational, 3 orientational and 1
conformational) have been used  to restrain the ligand in the bound conformation.
By using the DBC restraint, we were able to reduce the
number of needed potentials to one.}

\jerome{
DBC is the RMSD of ligand coordinates calculated in the frame of reference of the receptor's binding site. }\grace {It reflects the deviation of the ligand from its reference position after canceling the deviation of the binding site from its reference position.
This formulation allows the receptor-ligand assembly to be unrestrained in the simulation.}

\jerome{
Restraint forces applied to DBC are a function of the ligand atoms and the receptor atoms used to define the binding site.
However, because the dependency on receptor atoms only occurs through the roto-translational fit, DBC forces correspond to a rigid-body motion, and have no effect on receptor conformation.}

\jerome{Adjusting the width of the DBC restraint allows for adapting to different types of binding: well-defined poses can be narrowly surrounded by a tight flat well, whereas a broad ensemble of loosely bound configurations will require an equally broad restraint.
In all cases, the goal is to limit sampling throughout the decoupling process to those configurations that are relevant in the fully coupled state.}

\jerome{We shall now define the DBC coordinate more rigorously.}
Let $\vX_\proteintype$ be a $3n$-vector of coordinates of representative atoms of the binding site, and $\vX_\proteintype^\mathrm{ref}$ a set of fixed reference coordinates for those atoms.
This could encompass an entire macromolecule, or a binding domain, or a smaller region surrounding the site.
In the GPCR example of this work, those atoms are a set of alpha carbon atoms located near the superficial binding site; they need not surround it to define the site precisely.
We note $\bar\vx_\proteintype$ and $\bar\vx_\proteintype^\mathrm{ref}$ the respective centers of mass of those sets of coordinates.

\jerome{The receptor undergoes a combination of 3 types of motion: a global translation moving its center of mass by $ \bar\vx_\proteintype - \bar\vx_\proteintype^\mathrm{ref}$, a global rotation $\mathbf{R}^{-1}$ around its center of mass, and internal changes: conformational fluctuations or drift.
While we wish to preserve the conformational dynamics of the receptor, its global motion can be removed by applying the inverse rigid-body transformation to each atom $i$, yielding the roto-translated coordinates $\vx_i^{\prime}$:
\begin{eqnarray}
\vx_i^{\prime}&=& \mathbf{R} \left(\vx_i - \bar\vx_\proteintype\right) + \bar\vx_\proteintype^\mathrm{ref} ,
\end{eqnarray}
where the $\vx_i^{\prime}$ associated with the binding site are as close as possible to their reference locations because $\mathbf{R}$ is defined as minimizing the mean square deviation: 
\begin{equation} \sum_{k, \mathrm{rec}} \left(\vx_k^{\prime} - \vx_k^\mathrm{ref} \right)^2=\sum_{k, \mathrm{rec}} \left( \mathbf{R} \left(\vx_k - \bar\vx_\proteintype\right) + \bar\vx_\proteintype^\mathrm{ref}- \vx_k^\mathrm{ref} \right)^2
\end{equation}
}
\grace{
If the ligand has diffused in step with the protein complex, without changing its binding mode or conformation, its coordinates $\vx_l^{\prime}$ will be equal to their reference values.}

\grace{The distance to bound configuration collective variable $d$ reflects the deviation from that case. }  It is defined as the RMSD of the roto-translated ligand coordinates $\vx'_l$ with respect to their reference positions $\vx_l^\mathrm{ref}$: 
\begin{equation}  
d = \left[ \sum_{l, \mathrm{lig}} ( \vx_l^\prime - \vx_l^\mathrm{ref})^2 \right]^\frac{1}{2} = \left[ \sum_{l, \mathrm{lig}} \left(\mathbf{R} (\vx_l - \bar\vx_\proteintype) + \bar\vx_\proteintype^\mathrm{ref}-\vx_l^\mathrm{ref} \right)^2 \right]^\frac{1}{2}  \label{eq:ddef} 
\end{equation}

\grace{
$d$ can then be used to calculate a restraint potential $U_{\mathrm{DBC}}(d)$, such as a flat-well restraint potential as in Eq.~\ref{eq:fw}. The components of the restraint force on the ligand $\mathbf{F}^\mathrm{lig}_{\mathrm{DBC}}$ are calculated in the $x^{\prime}$ coordinate system (frame of reference of $\vX_\proteintype^\mathrm{ref}$), before $\mathbf{R}^{-1}$ is used to rotate the force vector back to the original coordinate system (current frame of reference of the simulation).
Conversely, a counter-force on receptor atoms arises from the dependence of the optimal translation and rotation on receptor coordinates, but this force acts globally and will not affect internal degrees of freedom.
}

Calculating the RMSD of coordinates that are roto-translated using a distinct group of atoms for the least-squares fit is a standard part of the Colvars Module toolkit,\cite{Fiorin2013} and was applied as is to implement the DBC coordinate.
An example script implementing the DBC coordinate and restraint in the NAMD Collective Variables Module is provided in Supplementary Information.

\subsection{Exclusion restraints}

In the AFEP decoupling simulation, ligand molecules other than the one being decoupled will tend to replace the disappearing bound ligand: this is all the more likely when the ligand is concentrated.
Formally, the end-point of the AFEP decoupling is a state where no ligand occupies the binding site, which in the present foramlism is defined by the range of DBC coordinate used in the bound ligand restraint.
Hence, to sample the unbound state, unperturbed ligands must be excluded from the binding site.
In many practical cases, the site forms a well-defined peak in the probability distribution of ligand positions, and thus a simple center-of-mass restraint will prevent binding without otherwise biasing the statistical distribution of unbound ligand molecules.

However, the case of the $\beta 2$-adrenergic receptor discussed below questions this assumption, as the superficial binding site is frequently lined with a second cholesterol molecule that transitions rapidly to the primary binding site if emptied, without apparent free energy barrier.
Preventing specifically this process without adding an arbitrary bias requires a very precise geometric definition of the binding mode, which is provided by the DBC.
We achieve this with a flat-well restraint that keeps the DBC above the bound threshold (the local minimum in the equilibrium DBC distribution).

While in theory this could be applied to all unbound ligands, in practice the least-squares fitting involved in the DBC restraint becomes computationally costly when iterated over a large number of molecules.
Thus ligands that start farther away from the site than they can diffuse over the timescale of a single simulation run may be left unrestrained.
For safety, an intermediate layer of ligands that are near but not immediately around the binding site may be prevented from binding by an inexpensive center-of-mass distance restraint.
The lists of ligand molecules affected by these various restraints can be updated periodically, for example between runs at each $\lambda$-value in the AFEP simulation.
This is the strategy followed by our NAMD implementation, for which a script is provided in SI.

\section{Application: Cholesterol binding to GPCRs}
Many crystal structures of G-protein coupled receptors (GPCRs) include resolved cholesterol, but the likelihood of these sites being occupied by cholesterol in a liquid membrane bilayer is unknown.  We selected three cholesterol binding sites on three different GCPRs, and applied this approach to predicting the cholesterol concentration required for 50\% occupancy when the GPCR is embedded in a POPC bilayer, with a total lipid to protein ratio of ~230 lipids (phospholipid or cholesterol) to 1 protein ($\ntotSUnit + \ntotLUnit = 230$).
\subsection{Methods}

\subsubsection{Receptor-bound Cholesterol System Setup} 

We used the crystal structures of the $\beta$2-adrenergic receptor ($\beta$2-AR, PDB id
3D4S), serotonin receptor type 2B (5-HT2B, 4NC3), and $\mu$-opioid receptor (5C1M) for setting up receptor-bound cholesterol systems.
A phosphatidylcholine (POPC) lipid-bilayer with 0.3 mole fraction of
cholesterol was generated using CHARMM-GUI Membrane Builder
\cite{Jo2008_CGUI,Brooks2009_Charmm,Jo2009}.
Each protein-cholesterol complex
was then embedded in the membrane by aligning the center of
the protein and the membrane along the z-direction and removing the overlapping
lipids. Residues were protonated according to their standard states at pH 7.4.
The systems were solvated using the \textsf{solvate} plugin of VMD  with TIP3P
water molecules, and Na+/Cl- ions were added to bring the system to a neutral
0.15 M concentration using the \textsf{autoionize} plugin.
The final system sizes were about 81$\times$80$\times$102~\AA$^3$ and included 162 POPC, 70 cholesterol and $\sim$ 12,700 water molecules, with a total of $\sim$70,000 atoms.

\subsubsection{Phospholipid-Cholesterol Mixed Bilayer System Setup}

For decoupling of a cholesterol molecule from the bulk of a bilayer with a
cholesterol mole fraction of $\cholfrac$ ranging from 0 to 40\% cholesterol, a mixed POPC/cholesterol bilayer was prepared using CHARMM-GUI Membrane Builder, as for
the protein-containing system. Na+/Cl- ions were added to the hydrated membrane to provide a neutral 0.15 M concentration. The final
systems 
contained $\sim$29,000 atoms including about 140 lipids and $\sim$5,400 water
molecules.  The bulk restraints ($^\brestraint$) enclosed one leaflet, hence half of the 140 lipids ($\ntotSUnit^{\brestraint}+ \ntotLUnit^{\brestraint} = 70$).
The GPCR-bilayer systems contain about 230 lipids, hence the bulk restraint volume factor is $\alpha = 70/230 = 0.3$.  Three AFEP replicas were run for each cholesterol fraction, and combined via linear averaging.  

\subsubsection{MD simulations}
The prepared systems were subjected to unrestrained molecular dynamics (MD)
simulations. MD simulations were performed using the NAMD\cite{Phillips2005}
2.12 simulation package updated with Colvars version 2017-09-18\cite{Fiorin2013},
CHARMM36 forcefield for the protein\cite{MacKerell1998,MacKerell2004a} and
phospholipids\cite{Klauda2010}, and modified CHARMM36 parameters\cite{Lim2012}
for cholesterol.

All MD simulations were performed using the NPT ensemble with weak coupling to
a Langevin thermostat and barostat at 300 K and 1 atm, respectively. Periodic
boundary conditions were employed, and the real space electrostatic
interactions were truncated at 12~\AA, while the long range components
were treated using PME method.\cite{Essman1995} Lennard-Jones interactions were
switched off smoothly between 10 and 12~\AA. All bonds to hydrogen
atoms were constrained using the SHAKE (non-water) or RATTLE (water)
algorithms. A multiple-time-step rRESPA method  was used, with fast
and slow time steps of 2 and 4~fs, respectively. 

All systems were first energy minimized for 10,000 steps and then equilibrated.
For the cholesterol-protein system, the position of protein C$\alpha$ atoms  and
heavy atoms of bound cholesterols were initially restrained using a 5
kcal/mol/\AA$^2$ harmonic force constant, which was gradually released
over 3~ns. After further unrestrained equilibration, a snapshot at 7~ns was
used as the initial structure for the free energy calculations to prevent
significant diffusion from the docked position. Unrestrained cholesterol-protein
MD simulations were then extended to 40~ns to study the dynamics of the
complex.  Similar unrestrained MD simulations were performed for cholesterol in
the bulk mixed membrane, and the results were used to determine boundaries for the
flat-bottom restraints in the free energy calculations. 

\subsubsection{Details of Restraints}

In the application presented here, involving cholesterol decoupled from a bulk that is a membrane bilayer, the membrane associated steps used a flat-well
restraint on the distance of the cholesterol center of mass (COM) from the monolayer midplane, as well as a conical
restraint on its orientation.  The total restraint potential $\Um$ for a single
cholesterol molecule in the membrane was the sum of the restraint
potential $\Uz$ on the COM position and $\Utheta$ on the orientation,

\begin{equation} 
    \Um = \Uz + \Utheta .
\end{equation}

In the coupled state, only the COM coordinate along the axis that is normal to
the membrane is limited; the COM is free to assume any position in the plane of
the membrane.  The variable used in the COM flat-well potential was the difference between this coordinate and
its average value in an unbiased equilibrium simulation, noted as $z$.  The potential for $\Uz$ used the form in Eq. \ref{eq:fw},
where the 
threshold distance $z_R$
is chosen to span the range
of $z$ values assumed by a cholesterol molecule in a single leaflet of the
bilayer, calculated in an unbiased equilibrium simulation. The restraint volume is $V^{\brestraint} = 2 z_{R} A^{\brestraint}$.  

Cholesterol molecules in the bulk of a lipid bilayer are expected to have a
strong preference for orientations parallel to the membrane normal, in which
the hydroxyl headgroup is exposed to solvent and the hydrophobic steroid rings
and hydrocarbon tail are buried in the core of the bilayer; unbiased
equilibrium simulations supported this expectation. The orientational restraint
$U_{\theta}$ imposed on the cholesterol in the membrane is therefore a function
of the azimuthal angle $\theta$ between a vector fit to the cholesterol
molecule (pointing toward the hydroxyl) and the normal axis, using the form in Eq. \ref{eq:fw}, 
where the threshold 
value, $\theta_R$, is half the opening angle of the conical restraint, chosen to
encompass the range of $\theta$ values assumed by a cholesterol molecule in a
single leaflet of the bilayer in an unbiased equilibrium simulation. In the decoupled state, imposing these restraints scales the phase-space volume by $\Omega_{\grestraint}/\Omega_{\brestraint}=1/({1 - \cos\theta_R}).$

Coarse protein-associated restraints confine the ligand to the general region of the protein binding site.  These restraints also used the form of Eq. \ref{eq:fw}, with the restraint variable $r$ equal to the difference between the ligand COM and its distance from its bound position,  and the threshold distance $r_{R}$ must be chosen to safely encompass all dispersion of the ligand COM in the bound state.  The volume of the restraint is $V^{\grestraint} = 4/3 \pi r_{R}^{3}$, so Eq. \ref{eq:bulkthree} becomes
\begin{equation}\frac{Z_\gphase^\grestraint} {Z_\gphase^\brestraint}= \frac{2\pi r_{R}^{3}} {3 z_{R} A^{\brestraint}}\frac{1}{1 - \cos\theta_R} \label{eq:theta_scale}\end{equation}

The DBC restraint involved 11 carbon atoms encompassing the four fused rings of cholesterol.
The moving frame of reference of the binding site was based on all alpha carbon atoms located within 15~\AA{} of the crystallographic cholesterol molecules (Supplementary Figure~1).
The Collective Variables Module\cite{Fiorin2013} was used to define the
restraints, with restraint coefficients in Table \ref{tbl:rest-values}.

\begin{table}[!ht]
\centering
\caption{{\bf Parameters used for restraint potentials.} Restraints applied to cholesterol molecule while decoupled from bulk or from protein binding site. All restraint potentials used the flat well potential in Eq. \ref{eq:fw}. COM refers to cholesterol COM.}
\label{tbl:rest-values}
\begin{tabular}{c|p{2.25in}cc}
  \hline
 Restraint  & Variable ($\xi$)  & Force constant  ($k_{\xi})$ & Threshold $(\xi_{\mathrm{max}})$  \\
  \hline
  \hline
   \multirow{4}{*}{bulk ($^{ \brestraint}$)}   & COM vertical distance from leaflet midplane &1000 kcal/mol.$\mbox{\AA}^2$ & $z_R = 11$   $\mbox{\AA}$\\
             &Angle between membrane normal and cholesterol axis    & 1000 kcal/mol.$\text{deg}^2$ &  $\theta_{R}= 0.14 \pi$ \\ 
             \hline
 coarse ($^{\grestraint}$) & COM Distance from crystallographic COM  &  100 kcal/mol.$\mbox{\AA}^2$ & $r_R = 5$~\AA   \\
 \hline
 DBC ($^{\prestraint}$) &  RMSD from crystallographic coordinates (Eq.~\ref{eq:ddef})  & 100 kcal/mol.\AA$^2$ & $d = 2$~\AA
\end{tabular}
\begin{flushleft}
\end{flushleft}
\end{table}

\subsubsection{Alchemical Free Energy Perturbation Calculations}
Decoupling via AFEP was carried using a total of 47 windows were spaced by
$\Delta\lambda = 0.05$ when $0 \leq \lambda < 0.5$,
$\Delta\lambda = 0.025$ until $\lambda = 0.8$,
$\Delta\lambda = 0.01$ until $\lambda = 0.95$, and
$\Delta\lambda = 0.005$ until $\lambda = 1$.
Simulations were performed
sequentially (i.e. the initial configuration for $\lambda_{i + 1}$  was the
final configuration from the run at $\lambda_i$).
Each window was run for 600~ps for equilibration purposes, followed by 2~ns of data collection.
Thus the total simulation time for a complete AFEP simulation was 122.2~ns.
We tested convergence of the cumulative average for each window by monitoring the
progression of  $\langle dG \rangle$ for each $\lambda$.  

To calculate $\zratFour$, RFEP simulations were run using the Colvars Module\cite{Fiorin2013} within NAMD, and 21
equally-spaced $\lambda$ values ranging from 1.0 (full restraints) to 0.0 (zero
restraints). Each $\lambda$ simulation was run for 400 ps, with the first 80ps 
discarded as equilibration. 
Restraint free energies were calculated from those simulations using the Simple Overlap Sampling estimator~\cite{LuN2003}.

\subsection{Results}
  
    \begin{figure}[!ht]
    \centering
    \includegraphics[width=6in]{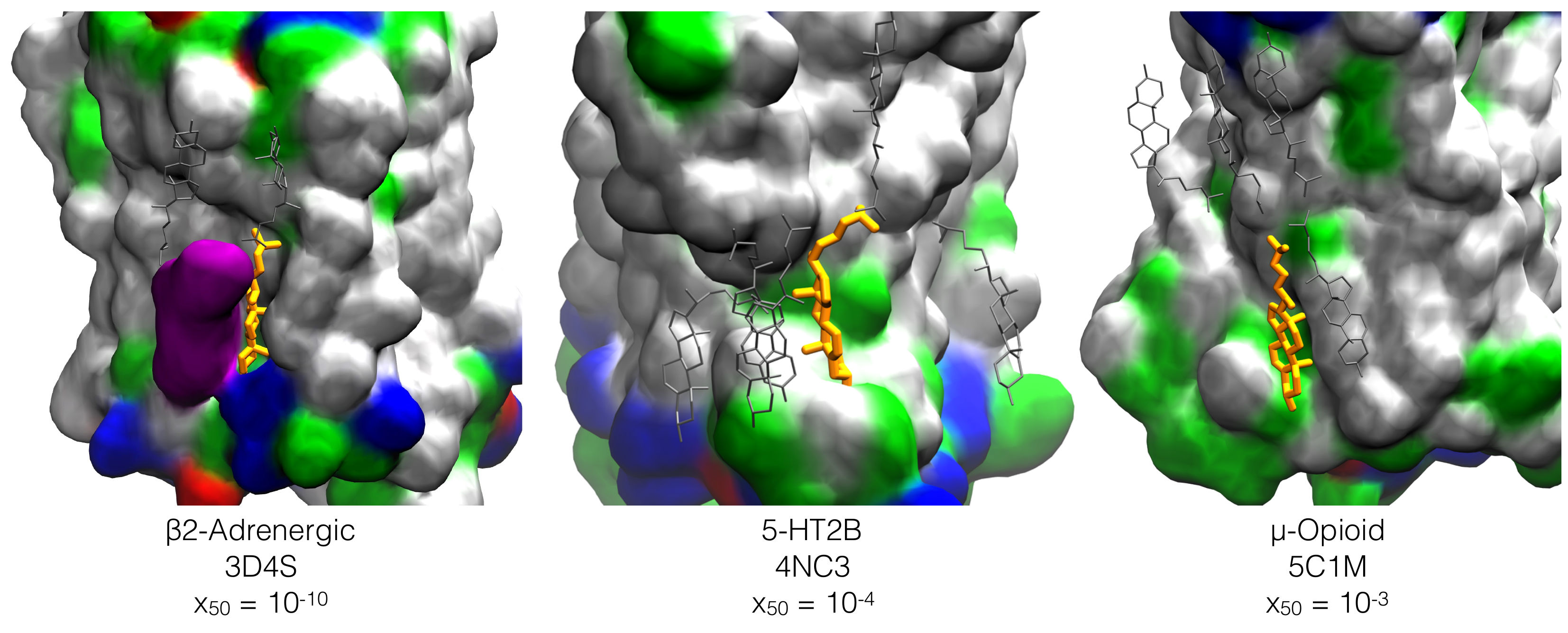}
    \caption{ {\bf Three crystallographic sites for cholesterol on three different GPCRs.} Protein is drawn in space-filling and colored by residue type : hydrophobic (white), polar (green), acidic (red), basic (blue). A second crystallographic cholesterol in the $\beta-2$adrenergic structure 3D4S is also shown in space-filling, colored purple.  A subset of the additional cholesterol molecules placed by CHARMM-GUI Membrane Builder for a mixture of 7:3 POPC:CHOL are in silver. Crystallographic cholesterol molecule in sites characterized in this work are in orange, and are residue 402,1203, and 404 in structure 3D4S, 4NC3, and 5C1M respectively. The corresponding half-saturation cholesterol fraction $x_{50}$ is from Table \ref{tab:protresults}; for the $\mu$-opioid receptor, $\focc < 50\%$ for the entire cholesterol range.  
   } 
    \label{fig:gpcrs}
  \end{figure}
  
\subsubsection{Non-ideality of Cholesterol In POPC Bilayers} 

\begin{figure}[!ht]
    \centering
    \includegraphics[scale=0.5]{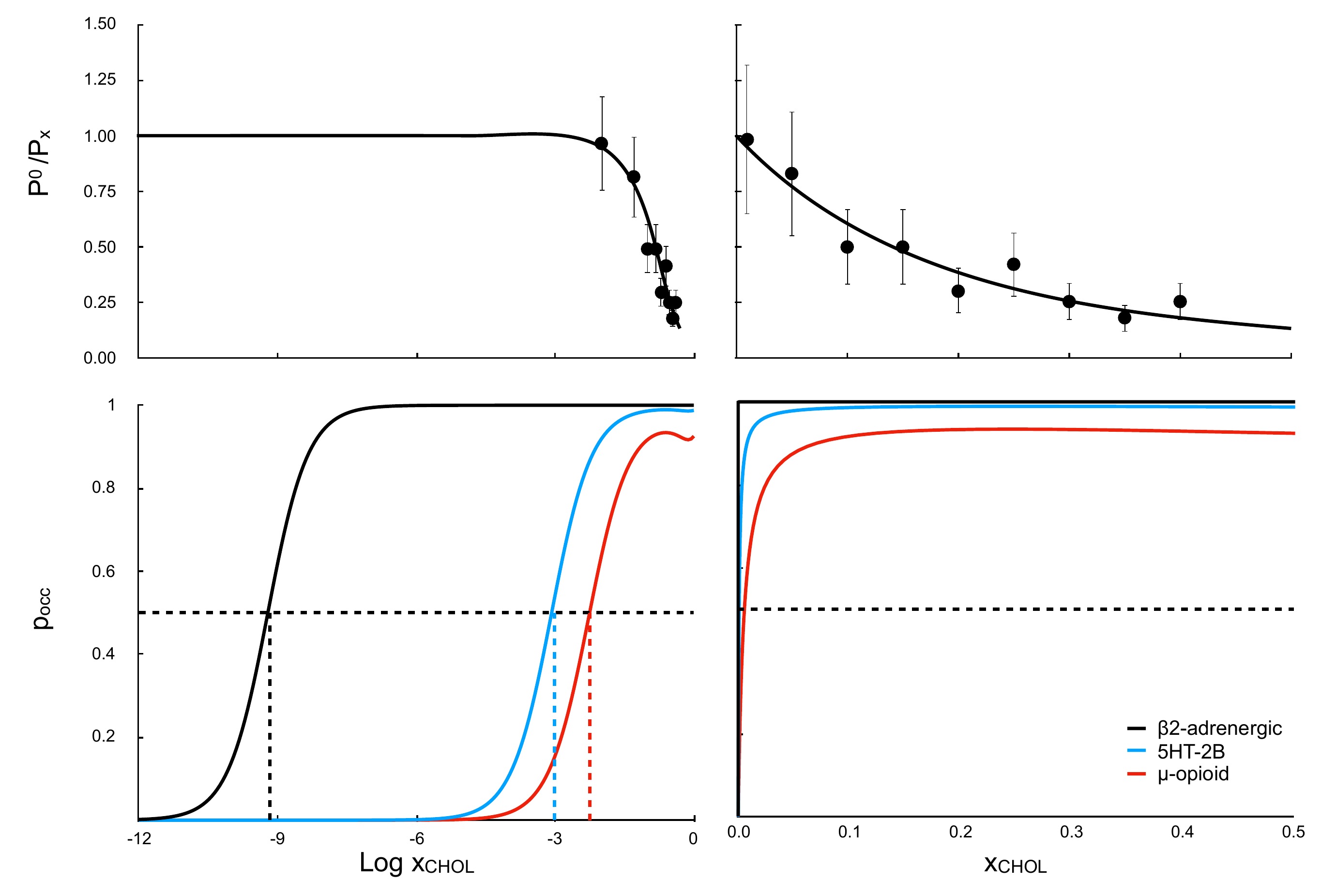}
    \caption{ {\bf Dependence of predicted site occupancy on ligand concentration, for cholesterol binding to crystallographic sites on three different GPCRs.}
    Top: Effect of cholesterol fraction on the normalized gas/bulk partition coefficient $\solvrat{o}{}^{0}/\solvrat{o}{\shortcholfrac}$, as calculated by AFEP decoupling of a single cholesterol molecule from a POPC:CHOL bilayer. Points represent average of three replicas, with standard error bars shown.  Curve represents $\frac{1}{\solvrat{o}{\shortcholfrac}}= \frac{1}{\solvrat{o}{}^{0}} \exp[h^{0}(1-\cholfrac)^{2}/RT]$ as in Eq.~\ref{eq:binary}, where $h^{0} = 1.6\pm0.6$ kcal/mol and $\solvrat{o}{}^{0} = 3\times10^{14}$, and both are free fit parameters. Reciprocals are plotted because they are directly used in the double-decoupling scheme. Bottom: Probability $\focc$ of occupied cholesterol-binding sites for three different GPCRs (Figure~\ref{fig:gpcrs}) in POPC:CHOL bilayers according to $\focc = (1 + \frac{1}{\probrat{\concL} \cholfrac})^{-1}$ and the values for $\probrat{\concL}$ in Table \ref{tab:protresults}. Left and right columns show the same data plotted vs $\log \cholfrac$ or $\cholfrac$, respectively.  Colored dashed lines indicate the midpoint cholesterol fraction $x_{50}$ for each receptor.  }    
    \label{fig:chol-frac}
  \end{figure}

$\solvrat{o}{\concL}$ was measured for decoupling a single cholesterol molecule from a POPC bilayer at multiple cholesterol concentrations ranging from trace to 40\%, via AFEP decoupling, with results shown in Fig.~\ref{fig:chol-frac}A and steps summarized in Table~\ref{tab:bulkresults}.  Non-ideality was confirmed, and $\solvrat{o}{\concL}$ decreased with increasing cholesterol concentration over most of this range.  The data was well-fit by the model in Eq.~\ref{eq:binary} ($P^{0} =  3\times10^{14}$ and $h^{0} = 1.6\pm0.6$kcal/mol), indicating that pair interactions between cholesterol and POPC were 1.6~kcal/mol less favorable, on average, than interaction between like lipids.
We extrapolate the fit to $\cholfrac = 0.5$, which is at the upper limit of cholesterol fractions in stable lipid bilayers.
Typical values for the phase boundary between mixed and liquid-ordered phases of binary mixtures of CHOL and POPC lie between 0.3-0.4 at room temperature.\cite{Marsh2010}

\subsubsection{Cholesterol Affinity for GPCRs}
Decoupling of cholesterol from three different GPCRs revealed significant differences in binding probability. Predicted concentration dependence of $\focc$ for cholesterol in crystallographic sites for three different GPCRs is shown in Fig.~\ref{fig:chol-frac}B, with values from AFEP recorded in Table~\ref{tab:protresults}.The $\beta$2-adrenergic receptor structure 3D4S\cite{Hanson2008} is one of the earliest high-resolution GPCR structures and the first containing resolved cholesterol. The predicted midpoint concentration $x_{50}$ is 1 cholesterol molecule per $10^{10}$ total lipid molecules. Occupation of this binding mode by cholesterol was sufficiently strong that toward the end of a decoupling run in 30\% cholesterol, a second cholesterol would reliably fill the site; this prompted the development of the exclusion restraints. 

\grace{Midpoint concentrations for the other two GPCRs tested, the 5-HT2 receptor in structure 4CN3 ($x_{50} = 0.01\%$ cholesterol, or 1 cholesterol molecule per $10^{4}$ lipid molecules) and the $\mu$-opioid receptor in structure 5C1M ($x_{50} = 0.1\%$ cholesterol or 1 cholesterol per $10^{3}$ lipid molecules) were significantly higher.} Gimpl has recently\cite{Gimpl2016} provided a comprehensive discussion of different binding modes of cholesterol for GPCRs; all three binding sites provide favorable hydrogen-bonding interactions for the hydroxyl and hydrophobic contacts for the tetracyclic ring system.
\grace{Another difference may explain the dramatically higher predicted occupancy of the 3D4S binding site: the presence of a second bound cholesterol molecule, which likely stabilizes the first cholesterol molecule through favorable stacking interactions of the two smooth faces. }

It is important to note that these trends may all be dependent upon the ``solvent'' lipid as well; cholesterol may have higher or lower probability binding to these sites in  bilayers composed of phospholipids with different head groups or saturation states. 

\grace{The fitted activity model based on Equation~\ref{eq:binary_chempot} predicts a non-monotonic variation of cholesterol chemical potential as a function of concentration, with a slight decrease above 12.5\%. For high-affinity sites that reach saturation at $\cholfrac << 12.5\%$, including the $\beta 2$-adrenergic and 5HT-2B receptor, the binding curve remains sigmoidal. 
In the real membrane, as the amount of cholesterol is much more than 10\%, the bulk becomes significantly more favorable, and the effective affinity of cholesterol for the protein site becomes increasingly weaker.  These competing effects result in a plateau for $\focc$ at less than 100\% occupancy for the $\mu$-opioid receptor.  An even lower affinity site would plateau at $\focc < 50\%$, making the notion of concentration at half-occupation irrelevant for low-affinity binding sites. }We note that casting our results for this receptor in the classic form of an affinity constant $K_A$ would not allow for accurate binding predictions for the typical range of cholesterol-phospholipid binary mixtures.

  \renewcommand{\arraystretch}{2.0}
  \begin{table}[!ht]
  \centering
    \caption{{Values for individual ratios of bulk-associated partition functions in Eq.~\ref{eq:zratio_simplified}, for cholesterol in a binary CHOL:POPC membrane.} \label{tab:bulkresults} 
}
    \begin{tabular}{cc|p{5cm}|p{6cm}}
      \hline
      Process & Ratio     &  Method     & Result                 \\
      \hline
      \hline
\makecell{apply bulk\\restraints} & $\zratOne$ & Eq.~\ref{eq:bulkone} with ~{$\alpha = 0.3$;}{~$\ntotLUnit = 230 \cholfrac$}
 & $10^{2} \cholfrac$ \\ 
\makecell{decouple\\from bulk}  & $\zratTwo$&  Fit of AFEP data for $\frac{1}{\solvrat{}{\shortcholfrac}}$ to Eq.~\ref{eq:binary}; see Figure~\ref{fig:chol-frac}A. &{$10^{-14}e^{2.7 (1-\cholfrac)^{2}}$} \\
\makecell{switch to\\isotropic restraint} & $\zratThree$ & Eq.~\ref{eq:theta_scale}~with $\theta_{R} = 0.14\pi, r_{R} =5 \mbox{\AA},  z_{R} = 11\mbox{\AA},  A^{\brestraint} = 3600\mbox{\AA}^{2}$&  $10^{-1}$\\ 
\hdashline
\makecell{bulk to\\gas phase} & $\frac{Z_\bphase^- Z_\gphase^\grestraint} {Z_\bphase^\norestraint}$& Product of previous 3 rows & {$10^{-13}\cholfrac e^{2.7 (1-\cholfrac)^{2}}$}\\ 
    \end{tabular}
  \end{table}

   \begin{table}[!ht]
  \centering
    \caption{ \label{tab:protresults}Values for individual ratios of protein-related partition functions in Eq.~\ref{eq:zratio_simplified}, for cholesterol binding to crystallographic sites on three GPCRs.   Predictions as functions of cholesterol fraction are plotted in Fig.~\ref{fig:chol-frac}B. AFEP values reflect the geometric mean of 3 replicas; see SI Table 1 for values from individual replicas and calculation of statistics.  See Table 5 for calculation of $\frac{ Z_\bphase^- Z_\gphase^\grestraint} {Z_\bphase^\norestraint}$. \\
\noindent  $^{*}$ restraints are designed so ${Z_\pphase^\occrestraint} \sim Z_\pphase^\prestraint$ and $Z_\gphase^\prestraint \sim Z_\gphase^{\grestraint\prestraint}$\\
}
    \begin{tabular}{cc|c|cccc}
      \hline
      Process & Ratio     &  Method    & $\beta$2-adrenergic    & 5-HT2B & $\mu$-opioid  \\
          \hline
add DBC restraint & $\frac{Z_\gphase^{\grestraint\prestraint}} {Z_\gphase^\grestraint}$&  RFEP  & $10^{-3}$ & $10^{-3}$&$10^{-3}$ \\
couple to receptor$^{}$ & $\frac{Z_\pphase^\prestraint} {Z_\gphase^\prestraint Z_\bphase^-}$& AFEP  &${10^{25}}$  &${10^{19}}$&${10^{18}}$ \\
\hdashline
bind from gas phase & $\frac{Z_\pphase^\occrestraint} {Z_\gphase^\grestraint Z_\bphase^-}$ & \makecell{Product of \\ previous 2 rows{$^*$}} &$10^{22}$ &$10^{16}$ &$10^{15}$\\
\hline\hline
\multirow{2}{*}{bind from bulk} & \multirow{2}{*}{$\probrat{{\shortcholfrac}}$} &\multirow{2}{*}{$\frac{ Z_\pphase^\occrestraint } { Z_\gphase^\grestraint Z_\bphase^- }\times\frac{ Z_\bphase^- Z_\gphase^\grestraint} {Z_\bphase^\norestraint}\frac{1}{\cholfrac}$}  & \multicolumn{3}{c}{${e^{2.7 (1-\cholfrac)^{2}}~~\times}$} \\
& && $10^{9}$  & $10^{3}$& $10^{2}$ \\ 
\hline
half-saturation & $x_{50}^{}$ & \makecell{$\cholfrac$ for which\\$\probrat{\shortcholfrac} \cholfrac= 1$} &  $10^{-9}$$^{}$ & $10^{-3}$$^{}$ & ${10^{-2}}$$^{}$   
   \end{tabular}
  \end{table}

\section{Discussion}

We have presented an approach for estimating probabilities that a binding site will be occupied by a ligand transferred from a complex, crowded environment such as a mixed lipid membrane. This approach represents an extension of classic AFEP methods
for transfer from dilute solution to a protein binding site, and consequently
requires no specification or sampling of binding or unbinding pathways.
We have provided a generalized derivation that can be applied to transfer
of the ligand from non-dilute concentrations in a complex fluid.  We have further provided a formalism for interpretation of results that emphasizes connection to laboratory observables in the concentration regime of interest.  This approach circumvents questions involving arbitrary standardized concentrations by allowing the receptor concentration (whatever it is) to serve as a natural concentration scale.

\grace{There are limited options for experimental assays to predict free energies for cholesterol binding to many transmembrane proteins, including GPCRs. Available methods introduce significant uncertainty in both the independent variable (local cholesterol concentration) and the dependent variable (estimations of binding site occupancy from functional response or structural changes). The infeasibility of an experimental affinity measurement for this process makes a rigorous affinity calculation both particularly relevant and nearly impossible to validate quantitatively. }

\grace{Nonetheless, there is experimental evidence for high affinity cholesterol binding sites for GPCRs.  Milon and colleagues \cite{Gater2014} measured the unfolding temperature of the $\beta$2-adrenergic receptor in the lipidic cubic phase as a function of cholesterol concentration,  finding effects on the folding temperature at the minimum concentration they tested, a cholesterol mol fraction of 0.2\%.  They estimated that this corresponded to a subnanomolar dissociation constant, but volume-based concentration scales for lipids are challenging to interpret, as discussed in the Introduction.  }

The double-decoupling technique inherent in the application of AFEP,
dictates that the free energy of the overall transfer process uses two primary
calculations, which quantify the free energy for transfer of the ligand from
either the bulk or protein phases to vacuum.  The validity of the AFEP
approach for each of these two calculations has been previously established
(see ref.~\citenum{Chodera2011} for a recent review and refs.~\citenum{LeBard2012,Gumbart2013,Wang2015} for recent applications); a primary
novelty in our approach lies in the consideration of more complex bulk phases and, to make this tractable, the use of simplified restraint schemes.

Potential computational sources of error, therefore, are those that need to be
considered in any AFEP calculation; namely, poor convergence.  In most AFEP
calculations of binding affinities, the use of well-designed restraints can
significantly improve convergence. We present here a set of restraints that improves convergence while minimizing the complexity of the overall thermodynamic cycle.  

The approach relies on the use of distance-to-bound configuration (DBC) restraints, which are effective at narrowly surrounding a well-defined binding pose.
This becomes a limitation if several binding poses are relevant to the target binding site and ligand.
Then, two alternate avenues may be used:
\begin{itemize}
 \item a broader flat well may be used, so that it encompasses all the relevant poses. This will increase the convergence time of the restrained AFEP simulation, potentially incurring a high cost in additional sampling;
 \item a separate set of simulations could be run for each identified binding pose, each one using narrow DBC restraints.
 For each pose, three simulations would be required: AFEP decoupling from the binding site under restraints, and free energy calculations of the DBC restraint free energy, for adding the restraint in vacuum and removing it in the binding site.
\end{itemize}

Compared to other restraint schemes, DBC restraints are able to delineate a superficial binding site as precisely as an internal one, as shown by the present GPCR example.
Classic approaches that define the center of the site as the center of a set or receptor atoms are only well-suited to the geometry of an internal binding cavity; they would not apply to such a case of superficial binding.

\grace{We suggested that the dramatically higher predicted occupancy of the $\beta_{2}$ adrenergic binding site could result from the presence of the second bound cholesterol molecule, which likely stabilizes the first cholesterol molecule through favorable stacking interactions of the two smooth faces. If so, removal of this second cholesterol molecule would reduce the measured receptor-coupling factor $\zratFive$, shifting the midpoint concentration for the first site to the right (probably significantly so).  This cooperativity provides an interesting example of a concentration regime (below the midpoint concentration of the second cholesterol molecule) in which the approximation that $\zratFive$ is independent of $\concL$ may not be reliable. It is possible that such favorable interactions may also underly the positive value of $h^0$, but only over concentrations for which cholesterol molecules typically only have one other cholesterol molecule as a nearest-neighbor.  }

Although the application here involved a membrane, \grace{the method should also be suitable for binding to proteins in non-dilute aqueous solution or a range of complex bulk phases, including detergent micelles, polymer melts, nematic or smectic liquid crystals, cubic or hexagonal lipid phases, or lipid domains within monolayers or bilayers. } The implementation does assume that receptors do not interact, and cannot be used to predict coupling between ligand concentration and protein organization, including effects of cholesterol concentration on GPCR dimerization.  They can still potentially provide insight into such processes; here, the results for the bulk and unitary systems suggest that segregation of cholesterol from phospholipids within protein grooves or behind other cholesterol molecules is highly favorable; dimerization of GPCRs around central cholesterol molecules (as investigated in~\cite{Prasanna2014}) segregates cholesterol molecules quite effectively, indicating a variant of the lipophobic effect~\cite{Mokrab2009,Duneau2013,Duneau2017} could be a potential driving force for oligomerization.

\begin{acknowledgement}
GB and RS were supported by research grants  NSF MCB1330728 and NIH P01GM55876-14A1. TJ was supported by an NIH 5T32GM112596-03.  
JH acknowledges funding from the French Agence Nationale de la Recherche through LABEX DYNAMO (ANR-11-LABX-0011). This project was supported with computational resources from the National Science Foundation XSEDE program through allocation NSF-MCB110149 as well as a local cluster funded by NSF-DBI1126052. 
\end{acknowledgement}

\begin{suppinfo}

  \paragraph{SI Figure 1.} Definition of a distance-to-bound-configuration coordinate.
  \paragraph{SI Figure 2.} Cartoon of rotations and translations in calculation of a distance-to-bound-configuration restraint force.
  \paragraph{SI Table 1.} Dispersion in protein decoupling $\left(\frac{Z_r^\triangle} {Z_g^\triangle Z_b^-}\right)$ calculations across replicas, and resulting uncertainty in $x_{50}$. 
  \paragraph{SI Text File 1.} NAMD Colvars configuration for DBC and center-of-mass restraints.
  \paragraph{SI Text File 2.} Analysis script for calculating the rows of Tables \ref{tab:bulkresults} and \ref{tab:protresults}, and the predictions in \ref{fig:chol-frac}.
  
   This information is available free of charge via the Internet at http://pubs.acs.org 
\end{suppinfo}

\bibliography{library}

\providecommand{\latin}[1]{#1}
\makeatletter
\providecommand{\doi}
  {\begingroup\let\do\@makeother\dospecials
  \catcode`\{=1 \catcode`\}=2 \doi@aux}
\providecommand{\doi@aux}[1]{\endgroup\texttt{#1}}
\makeatother
\providecommand*\mcitethebibliography{\thebibliography}
\csname @ifundefined\endcsname{endmcitethebibliography}
  {\let\endmcitethebibliography\endthebibliography}{}
\begin{mcitethebibliography}{55}
\providecommand*\natexlab[1]{#1}
\providecommand*\mciteSetBstSublistMode[1]{}
\providecommand*\mciteSetBstMaxWidthForm[2]{}
\providecommand*\mciteBstWouldAddEndPuncttrue
  {\def\EndOfBibitem{\unskip.}}
\providecommand*\mciteBstWouldAddEndPunctfalse
  {\let\EndOfBibitem\relax}
\providecommand*\mciteSetBstMidEndSepPunct[3]{}
\providecommand*\mciteSetBstSublistLabelBeginEnd[3]{}
\providecommand*\EndOfBibitem{}
\mciteSetBstSublistMode{f}
\mciteSetBstMaxWidthForm{subitem}{(\alph{mcitesubitemcount})}
\mciteSetBstSublistLabelBeginEnd
  {\mcitemaxwidthsubitemform\space}
  {\relax}
  {\relax}

\bibitem[Malde \latin{et~al.}(2011)Malde, Zuo, Breeze, Stroet, Poger, Nair,
  Oostenbrink, and Mark]{Malde2011}
Malde,~A.~K.; Zuo,~L.; Breeze,~M.; Stroet,~M.; Poger,~D.; Nair,~P.~C.;
  Oostenbrink,~C.; Mark,~A.~E. An Automated Force Field Topology Builder (ATB)
  and Repository: Version 1.0. \emph{Journal of chemical theory and
  computation} \textbf{2011}, \emph{7}, 4026--4037\relax
\mciteBstWouldAddEndPuncttrue
\mciteSetBstMidEndSepPunct{\mcitedefaultmidpunct}
{\mcitedefaultendpunct}{\mcitedefaultseppunct}\relax
\EndOfBibitem
\bibitem[Vanommeslaeghe and MacKerell(2012)Vanommeslaeghe, and
  MacKerell]{Vanommeslaeghe2012}
Vanommeslaeghe,~K.; MacKerell,~A.~D. {Automation of the CHARMM General Force
  Field (CGenFF) I: bond perception and atom typing.} \emph{Journal of chemical
  information and modeling} \textbf{2012}, \emph{52}, 3144--54\relax
\mciteBstWouldAddEndPuncttrue
\mciteSetBstMidEndSepPunct{\mcitedefaultmidpunct}
{\mcitedefaultendpunct}{\mcitedefaultseppunct}\relax
\EndOfBibitem
\bibitem[Vanommeslaeghe \latin{et~al.}(2012)Vanommeslaeghe, Raman, and
  MacKerell]{Vanommeslaeghe2012a}
Vanommeslaeghe,~K.; Raman,~E.~P.; MacKerell,~A.~D. {Automation of the CHARMM
  General Force Field (CGenFF) II: assignment of bonded parameters and partial
  atomic charges.} \emph{Journal of chemical information and modeling}
  \textbf{2012}, \emph{52}, 3155--68\relax
\mciteBstWouldAddEndPuncttrue
\mciteSetBstMidEndSepPunct{\mcitedefaultmidpunct}
{\mcitedefaultendpunct}{\mcitedefaultseppunct}\relax
\EndOfBibitem
\bibitem[Mayne \latin{et~al.}(2013)Mayne, Saam, Schulten, Tajkhorshid, and
  Gumbart]{Mayne2013}
Mayne,~C.~G.; Saam,~J.; Schulten,~K.; Tajkhorshid,~E.; Gumbart,~J.~C. {Rapid
  parameterization of small molecules using the Force Field Toolkit.}
  \emph{Journal of computational chemistry} \textbf{2013}, \emph{34},
  2757--70\relax
\mciteBstWouldAddEndPuncttrue
\mciteSetBstMidEndSepPunct{\mcitedefaultmidpunct}
{\mcitedefaultendpunct}{\mcitedefaultseppunct}\relax
\EndOfBibitem
\bibitem[Lundborg and Lindahl(2015)Lundborg, and Lindahl]{Lundborg2015}
Lundborg,~M.; Lindahl,~E. Automatic GROMACS topology generation and comparisons
  of force fields for solvation free energy calculations. \emph{The journal of
  physical chemistry. B} \textbf{2015}, \emph{119}, 810--823\relax
\mciteBstWouldAddEndPuncttrue
\mciteSetBstMidEndSepPunct{\mcitedefaultmidpunct}
{\mcitedefaultendpunct}{\mcitedefaultseppunct}\relax
\EndOfBibitem
\bibitem[Zheng \latin{et~al.}(2016)Zheng, Tang, He, Du, Xu, Wang, Xu, and
  Lin]{Zheng2016}
Zheng,~S.; Tang,~Q.; He,~J.; Du,~S.; Xu,~S.; Wang,~C.; Xu,~Y.; Lin,~F. VFFDT: A
  New Software for Preparing AMBER Force Field Parameters for Metal-Containing
  Molecular Systems. \emph{Journal of chemical information and modeling}
  \textbf{2016}, \emph{56}, 811--818\relax
\mciteBstWouldAddEndPuncttrue
\mciteSetBstMidEndSepPunct{\mcitedefaultmidpunct}
{\mcitedefaultendpunct}{\mcitedefaultseppunct}\relax
\EndOfBibitem
\bibitem[Dodda \latin{et~al.}(2017)Dodda, Cabeza~de Vaca, Tirado-Rives, and
  Jorgensen]{Dodda2017}
Dodda,~L.~S.; Cabeza~de Vaca,~I.; Tirado-Rives,~J.; Jorgensen,~W.~L. LigParGen
  web server: an automatic OPLS-AA parameter generator for organic ligands.
  \emph{Nucleic acids research} \textbf{2017}, \emph{45}, W331--W336\relax
\mciteBstWouldAddEndPuncttrue
\mciteSetBstMidEndSepPunct{\mcitedefaultmidpunct}
{\mcitedefaultendpunct}{\mcitedefaultseppunct}\relax
\EndOfBibitem
\bibitem[Woo and Roux(2005)Woo, and Roux]{Woo2005a}
Woo,~H.-j.; Roux,~B. {Calculation of absolute protein-ligand binding free
  energy from computer simulations.} \emph{Proceedings of the National Academy
  of Sciences of the United States of America} \textbf{2005}, \emph{102},
  6825--30\relax
\mciteBstWouldAddEndPuncttrue
\mciteSetBstMidEndSepPunct{\mcitedefaultmidpunct}
{\mcitedefaultendpunct}{\mcitedefaultseppunct}\relax
\EndOfBibitem
\bibitem[Gilson and Zhou(2007)Gilson, and Zhou]{Gilson2007}
Gilson,~M.~K.; Zhou,~H.-X. Calculation of Protein-Ligand Binding Affinities.
  \emph{Annual Review of Biophysics and Biomolecular Structure} \textbf{2007},
  \emph{36}, 21--42\relax
\mciteBstWouldAddEndPuncttrue
\mciteSetBstMidEndSepPunct{\mcitedefaultmidpunct}
{\mcitedefaultendpunct}{\mcitedefaultseppunct}\relax
\EndOfBibitem
\bibitem[Chipot(2008)]{Chipot2008}
Chipot,~C. \emph{Methods in Molecular Biology}; Springer Science Business
  Media, 2008; pp 121--144\relax
\mciteBstWouldAddEndPuncttrue
\mciteSetBstMidEndSepPunct{\mcitedefaultmidpunct}
{\mcitedefaultendpunct}{\mcitedefaultseppunct}\relax
\EndOfBibitem
\bibitem[Pohorille \latin{et~al.}(2010)Pohorille, Jarzynski, and
  Chipot]{Pohorille2010}
Pohorille,~A.; Jarzynski,~C.; Chipot,~C. {Good practices in free-energy
  calculations.} \emph{The journal of physical chemistry. B} \textbf{2010},
  \emph{114}, 10235--10253\relax
\mciteBstWouldAddEndPuncttrue
\mciteSetBstMidEndSepPunct{\mcitedefaultmidpunct}
{\mcitedefaultendpunct}{\mcitedefaultseppunct}\relax
\EndOfBibitem
\bibitem[H\'{e}nin \latin{et~al.}(2010)H\'{e}nin, Brannigan, Dailey, Eckenhoff,
  and Klein]{Henin2010b}
H\'{e}nin,~J.; Brannigan,~G.; Dailey,~W.~P.; Eckenhoff,~R.; Klein,~M.~L. {An
  atomistic model for simulations of the general anesthetic isoflurane.}
  \emph{The journal of physical chemistry. B} \textbf{2010}, \emph{114},
  604--12\relax
\mciteBstWouldAddEndPuncttrue
\mciteSetBstMidEndSepPunct{\mcitedefaultmidpunct}
{\mcitedefaultendpunct}{\mcitedefaultseppunct}\relax
\EndOfBibitem
\bibitem[Gumbart \latin{et~al.}(2013)Gumbart, Roux, and Chipot]{Gumbart2013}
Gumbart,~J.~C.; Roux,~B.; Chipot,~C. {Standard binding free energies from
  computer simulations: What is the best strategy?} \emph{Journal of chemical
  theory and computation} \textbf{2013}, \emph{9}, 794--802\relax
\mciteBstWouldAddEndPuncttrue
\mciteSetBstMidEndSepPunct{\mcitedefaultmidpunct}
{\mcitedefaultendpunct}{\mcitedefaultseppunct}\relax
\EndOfBibitem
\bibitem[Gilson \latin{et~al.}(1997)Gilson, Given, Bush, and
  McCammon]{Gilson1997a}
Gilson,~M.~K.; Given,~J.~A.; Bush,~B.~L.; McCammon,~J.~A. {The
  statistical-thermodynamic basis for computation of binding affinities: a
  critical review.} \emph{Biophysical journal} \textbf{1997}, \emph{72},
  1047--69\relax
\mciteBstWouldAddEndPuncttrue
\mciteSetBstMidEndSepPunct{\mcitedefaultmidpunct}
{\mcitedefaultendpunct}{\mcitedefaultseppunct}\relax
\EndOfBibitem
\bibitem[Boresch \latin{et~al.}(2003)Boresch, Tettinger, Leitgeb, and
  Karplus]{Boresch2003}
Boresch,~S.; Tettinger,~F.; Leitgeb,~M.; Karplus,~M. {Absolute Binding Free
  Energies: A Quantitative Approach for Their Calculation}. \emph{The Journal
  of Physical Chemistry B} \textbf{2003}, \emph{107}, 9535--9551\relax
\mciteBstWouldAddEndPuncttrue
\mciteSetBstMidEndSepPunct{\mcitedefaultmidpunct}
{\mcitedefaultendpunct}{\mcitedefaultseppunct}\relax
\EndOfBibitem
\bibitem[Wang \latin{et~al.}(2006)Wang, Deng, and Roux]{Wang2006}
Wang,~J.; Deng,~Y.; Roux,~B. {Absolute binding free energy calculations using
  molecular dynamics simulations with restraining potentials.}
  \emph{Biophysical journal} \textbf{2006}, \emph{91}, 2798--2814\relax
\mciteBstWouldAddEndPuncttrue
\mciteSetBstMidEndSepPunct{\mcitedefaultmidpunct}
{\mcitedefaultendpunct}{\mcitedefaultseppunct}\relax
\EndOfBibitem
\bibitem[Chen \latin{et~al.}(2015)Chen, Zhou, Wang, Zheng, Gao, and
  Zhou]{Chen2015}
Chen,~H.; Zhou,~X.; Wang,~A.; Zheng,~Y.; Gao,~Y.; Zhou,~J. {Evolutions in
  fragment-based drug design: the deconstruction-reconstruction approach}.
  \emph{Drug Discovery Today} \textbf{2015}, \emph{20}, 105--113\relax
\mciteBstWouldAddEndPuncttrue
\mciteSetBstMidEndSepPunct{\mcitedefaultmidpunct}
{\mcitedefaultendpunct}{\mcitedefaultseppunct}\relax
\EndOfBibitem
\bibitem[Zhang and Lazaridis(2006)Zhang, and Lazaridis]{Zhang2006}
Zhang,~J.; Lazaridis,~T. {Calculating the free energy of association of
  transmembrane helices.} \emph{Biophysical journal} \textbf{2006}, \emph{91},
  1710--23\relax
\mciteBstWouldAddEndPuncttrue
\mciteSetBstMidEndSepPunct{\mcitedefaultmidpunct}
{\mcitedefaultendpunct}{\mcitedefaultseppunct}\relax
\EndOfBibitem
\bibitem[Hanson \latin{et~al.}(2008)Hanson, Cherezov, Griffith, Roth, Jaakola,
  Chien, Velasquez, Kuhn, and Stevens]{Hanson2008}
Hanson,~M.~A.; Cherezov,~V.; Griffith,~M.~T.; Roth,~C.~B.; Jaakola,~V.~P.;
  Chien,~E.~Y.; Velasquez,~J.; Kuhn,~P.; Stevens,~R.~C. {A Specific Cholesterol
  Binding Site Is Established by the 2.8 {\AA} Structure of the Human
  $\beta$2-Adrenergic Receptor}. \emph{Structure} \textbf{2008}, \emph{16},
  897--905\relax
\mciteBstWouldAddEndPuncttrue
\mciteSetBstMidEndSepPunct{\mcitedefaultmidpunct}
{\mcitedefaultendpunct}{\mcitedefaultseppunct}\relax
\EndOfBibitem
\bibitem[Oates and Watts(2011)Oates, and Watts]{Oates2011}
Oates,~J.; Watts,~A. {Uncovering the intimate relationship between lipids,
  cholesterol and GPCR activation}. \emph{Current Opinion in Structural
  Biology} \textbf{2011}, \emph{21}, 802--807\relax
\mciteBstWouldAddEndPuncttrue
\mciteSetBstMidEndSepPunct{\mcitedefaultmidpunct}
{\mcitedefaultendpunct}{\mcitedefaultseppunct}\relax
\EndOfBibitem
\bibitem[Venkatakrishnan \latin{et~al.}(2013)Venkatakrishnan, Deupi, Lebon,
  Tate, Schertler, and {Madan Babu}]{Venkatakrishnan2013}
Venkatakrishnan,~A.~J.; Deupi,~X.; Lebon,~G.; Tate,~C.~G.; Schertler,~G.~F.;
  {Madan Babu},~M. {Molecular signatures of G-protein-coupled receptors}.
  \emph{Nature} \textbf{2013}, \emph{494}, 185--194\relax
\mciteBstWouldAddEndPuncttrue
\mciteSetBstMidEndSepPunct{\mcitedefaultmidpunct}
{\mcitedefaultendpunct}{\mcitedefaultseppunct}\relax
\EndOfBibitem
\bibitem[Gimpl(2016)]{Gimpl2016}
Gimpl,~G. {Interaction of G protein coupled receptors and cholesterol}.
  \emph{Chemistry and Physics of Lipids} \textbf{2016}, \emph{199},
  61--73\relax
\mciteBstWouldAddEndPuncttrue
\mciteSetBstMidEndSepPunct{\mcitedefaultmidpunct}
{\mcitedefaultendpunct}{\mcitedefaultseppunct}\relax
\EndOfBibitem
\bibitem[Grossfield(2011)]{Grossfield2011}
Grossfield,~A. {Recent progress in the study of G protein-coupled receptors
  with molecular dynamics computer simulations}. \emph{Biochimica et Biophysica
  Acta (BBA) - Biomembranes} \textbf{2011}, \emph{1808}, 1868--1878\relax
\mciteBstWouldAddEndPuncttrue
\mciteSetBstMidEndSepPunct{\mcitedefaultmidpunct}
{\mcitedefaultendpunct}{\mcitedefaultseppunct}\relax
\EndOfBibitem
\bibitem[Lee and Lyman(2012)Lee, and Lyman]{Lee2012}
Lee,~J.~Y.; Lyman,~E. {Predictions for cholesterol interaction sites on the
  A2Aadenosine receptor}. \emph{Journal of the American Chemical Society}
  \textbf{2012}, \emph{134}, 16512--16515\relax
\mciteBstWouldAddEndPuncttrue
\mciteSetBstMidEndSepPunct{\mcitedefaultmidpunct}
{\mcitedefaultendpunct}{\mcitedefaultseppunct}\relax
\EndOfBibitem
\bibitem[Sengupta and Chattopadhyay(2012)Sengupta, and
  Chattopadhyay]{Sengupta2012}
Sengupta,~D.; Chattopadhyay,~A. {Identification of Cholesterol Binding Sites in
  the Serotonin 1A Receptor}. \emph{The Journal of Physical Chemistry B}
  \textbf{2012}, \emph{116}, 12991--12996\relax
\mciteBstWouldAddEndPuncttrue
\mciteSetBstMidEndSepPunct{\mcitedefaultmidpunct}
{\mcitedefaultendpunct}{\mcitedefaultseppunct}\relax
\EndOfBibitem
\bibitem[Prasanna \latin{et~al.}(2014)Prasanna, Chattopadhyay, and
  Sengupta]{Prasanna2014}
Prasanna,~X.; Chattopadhyay,~A.; Sengupta,~D. {Cholesterol modulates the dimer
  interface of the $\beta$2- adrenergic receptor via cholesterol occupancy
  sites}. \emph{Biophysical Journal} \textbf{2014}, \emph{106},
  1290--1300\relax
\mciteBstWouldAddEndPuncttrue
\mciteSetBstMidEndSepPunct{\mcitedefaultmidpunct}
{\mcitedefaultendpunct}{\mcitedefaultseppunct}\relax
\EndOfBibitem
\bibitem[Sengupta and Chattopadhyay(2015)Sengupta, and
  Chattopadhyay]{Sengupta2015}
Sengupta,~D.; Chattopadhyay,~A. {Molecular dynamics simulations of
  GPCR–cholesterol interaction: An emerging paradigm}. \emph{BBA -
  Biomembranes} \textbf{2015}, \emph{1848}, 1775--1782\relax
\mciteBstWouldAddEndPuncttrue
\mciteSetBstMidEndSepPunct{\mcitedefaultmidpunct}
{\mcitedefaultendpunct}{\mcitedefaultseppunct}\relax
\EndOfBibitem
\bibitem[Phillips \latin{et~al.}(2012)Phillips, Kondev, Theriot, and
  Garcia]{Phillips2012}
Phillips,~R.; Kondev,~J.; Theriot,~J.; Garcia,~H. \emph{{Physical Biology of
  the Cell}}, 2nd ed.; Garland Sciences, 2012\relax
\mciteBstWouldAddEndPuncttrue
\mciteSetBstMidEndSepPunct{\mcitedefaultmidpunct}
{\mcitedefaultendpunct}{\mcitedefaultseppunct}\relax
\EndOfBibitem
\bibitem[Lu \latin{et~al.}(2003)Lu, Singh, and Kofke]{LuN2003}
Lu,~N.; Singh,~J.~K.; Kofke,~D.~A. Appropriate methods to combine forward and
  reverse free-energy perturbation averages. \emph{J. Chem. Phys.}
  \textbf{2003}, \emph{118}, 2977--2984\relax
\mciteBstWouldAddEndPuncttrue
\mciteSetBstMidEndSepPunct{\mcitedefaultmidpunct}
{\mcitedefaultendpunct}{\mcitedefaultseppunct}\relax
\EndOfBibitem
\bibitem[Rowlinson and Swinton(1982)Rowlinson, and Swinton]{Rowlinson1982}
Rowlinson,~J.~S.; Swinton,~F.~L. In \emph{Liquids and Liquid Mixtures:
  Butterworths Monographs in Chemistry}; Baldwin,~J.~E., Buckingham,~A.~D.,
  Danishefsky,~S., Eds.; Elsevier, 1982\relax
\mciteBstWouldAddEndPuncttrue
\mciteSetBstMidEndSepPunct{\mcitedefaultmidpunct}
{\mcitedefaultendpunct}{\mcitedefaultseppunct}\relax
\EndOfBibitem
\bibitem[Silver(1985)]{Silver1985}
Silver,~B. \emph{The Physical Chemistry of Membranes: An Introduction to the
  Structure and Dynamics of Biological Membranes}; Allen \& Unwin, 1985\relax
\mciteBstWouldAddEndPuncttrue
\mciteSetBstMidEndSepPunct{\mcitedefaultmidpunct}
{\mcitedefaultendpunct}{\mcitedefaultseppunct}\relax
\EndOfBibitem
\bibitem[Morris(2008)]{Morris2008}
Morris,~J.~W. \emph{Notes on the Thermodynamics of Solids (Chapter 18:
  Solutions)}; Department of Materials Science and Engineering, University of
  California-Berkeley, 2008\relax
\mciteBstWouldAddEndPuncttrue
\mciteSetBstMidEndSepPunct{\mcitedefaultmidpunct}
{\mcitedefaultendpunct}{\mcitedefaultseppunct}\relax
\EndOfBibitem
\bibitem[Brannigan and Brown(2005)Brannigan, and Brown]{Brannigan2005b}
Brannigan,~G.; Brown,~F. L.~H. {Composition dependence of bilayer elasticity.}
  \emph{J. Chem. Phys.} \textbf{2005}, \emph{122}, 074905\relax
\mciteBstWouldAddEndPuncttrue
\mciteSetBstMidEndSepPunct{\mcitedefaultmidpunct}
{\mcitedefaultendpunct}{\mcitedefaultseppunct}\relax
\EndOfBibitem
\bibitem[Woll \latin{et~al.}(2016)Woll, Murlidaran, Pinch, H{\'{e}}nin, Wang,
  Salari, Covarrubias, Dailey, Brannigan, Garcia, and Eckenhoff]{Woll2016b}
Woll,~K.~A.; Murlidaran,~S.; Pinch,~B.~J.; H{\'{e}}nin,~J.; Wang,~X.;
  Salari,~R.; Covarrubias,~M.; Dailey,~W.~P.; Brannigan,~G.; Garcia,~B.~A.;
  Eckenhoff,~R.~G. A Novel Bifunctional Alkylphenol Anesthetic Allows
  Characterization of GABAA Receptor Subunit Binding Selectivity in
  Synaptosomes. \emph{J Biol Chem} \textbf{2016}, 10.1074/jbc.M116.736975\relax
\mciteBstWouldAddEndPuncttrue
\mciteSetBstMidEndSepPunct{\mcitedefaultmidpunct}
{\mcitedefaultendpunct}{\mcitedefaultseppunct}\relax
\EndOfBibitem
\bibitem[LeBard \latin{et~al.}(2012)LeBard, H\'{e}nin, Eckenhoff, Klein, and
  Brannigan]{LeBard2012}
LeBard,~D.~N.; H\'{e}nin,~J.; Eckenhoff,~R.~G.; Klein,~M.~L.; Brannigan,~G.
  {General anesthetics predicted to block the GLIC pore with micromolar
  affinity.} \emph{PLoS computational biology} \textbf{2012}, \emph{8},
  e1002532\relax
\mciteBstWouldAddEndPuncttrue
\mciteSetBstMidEndSepPunct{\mcitedefaultmidpunct}
{\mcitedefaultendpunct}{\mcitedefaultseppunct}\relax
\EndOfBibitem
\bibitem[Ge and Roux(2010)Ge, and Roux]{Ge2010}
Ge,~X.; Roux,~B. {Absolute binding free energy calculations of sparsomycin
  analogs to the bacterial ribosome.} \emph{The journal of physical chemistry.
  B} \textbf{2010}, \emph{114}, 9525--39\relax
\mciteBstWouldAddEndPuncttrue
\mciteSetBstMidEndSepPunct{\mcitedefaultmidpunct}
{\mcitedefaultendpunct}{\mcitedefaultseppunct}\relax
\EndOfBibitem
\bibitem[Mobley \latin{et~al.}(2006)Mobley, Chodera, and Dill]{Mobley2006}
Mobley,~D.~L.; Chodera,~J.~D.; Dill,~K.~A. {On the use of orientational
  restraints and symmetry corrections in alchemical free energy calculations.}
  \emph{The Journal of chemical physics} \textbf{2006}, \emph{125},
  084902\relax
\mciteBstWouldAddEndPuncttrue
\mciteSetBstMidEndSepPunct{\mcitedefaultmidpunct}
{\mcitedefaultendpunct}{\mcitedefaultseppunct}\relax
\EndOfBibitem
\bibitem[Fiorin \latin{et~al.}(2013)Fiorin, Klein, and H\'{e}nin]{Fiorin2013}
Fiorin,~G.; Klein,~M.~L.; H\'{e}nin,~J. {Using collective variables to drive
  molecular dynamics simulations}. \emph{Molecular Physics} \textbf{2013},
  \emph{111}, 3345--3362\relax
\mciteBstWouldAddEndPuncttrue
\mciteSetBstMidEndSepPunct{\mcitedefaultmidpunct}
{\mcitedefaultendpunct}{\mcitedefaultseppunct}\relax
\EndOfBibitem
\bibitem[Jo \latin{et~al.}(2008)Jo, Kim, Iyer, and Im]{Jo2008_CGUI}
Jo,~S.; Kim,~T.; Iyer,~V.~G.; Im,~W. {CHARMM-GUI: A web-based graphical user
  interface for CHARMM}. \emph{Journal of Computational Chemistry}
  \textbf{2008}, \emph{29}, 1859--1865\relax
\mciteBstWouldAddEndPuncttrue
\mciteSetBstMidEndSepPunct{\mcitedefaultmidpunct}
{\mcitedefaultendpunct}{\mcitedefaultseppunct}\relax
\EndOfBibitem
\bibitem[Brooks \latin{et~al.}(2009)Brooks, Brooks, Mackerell, Nilsson,
  Petrella, Roux, Won, Archontis, Bartels, Boresch, Caflisch, Caves, Cui,
  Dinner, Feig, Fischer, Gao, Hodoscek, Im, Kuczera, Lazaridis, Ma,
  Ovchinnikov, Paci, Pastor, Post, Pu, Schaefer, Tidor, Venable, Woodcock, Wu,
  Yang, York, and Karplus]{Brooks2009_Charmm}
Brooks,~B.~R. \latin{et~al.}  {CHARMM: The biomolecular simulation program}.
  \emph{Journal of Computational Chemistry} \textbf{2009}, \emph{30},
  1545--1614\relax
\mciteBstWouldAddEndPuncttrue
\mciteSetBstMidEndSepPunct{\mcitedefaultmidpunct}
{\mcitedefaultendpunct}{\mcitedefaultseppunct}\relax
\EndOfBibitem
\bibitem[Jo \latin{et~al.}(2009)Jo, Lim, Klauda, and Im]{Jo2009}
Jo,~S.; Lim,~J.~B.; Klauda,~J.~B.; Im,~W. {CHARMM-GUI Membrane Builder for
  mixed bilayers and its application to yeast membranes.} \emph{Biophys. J.}
  \textbf{2009}, \emph{97}, 50--58\relax
\mciteBstWouldAddEndPuncttrue
\mciteSetBstMidEndSepPunct{\mcitedefaultmidpunct}
{\mcitedefaultendpunct}{\mcitedefaultseppunct}\relax
\EndOfBibitem
\bibitem[Phillips \latin{et~al.}(2005)Phillips, Braun, Wang, Gumbart,
  Tajkhorshid, Villa, Chipot, Skeel, Kal\'{e}, and Schulten]{Phillips2005}
Phillips,~J.~C.; Braun,~R.; Wang,~W.; Gumbart,~J.; Tajkhorshid,~E.; Villa,~E.;
  Chipot,~C.; Skeel,~R.~D.; Kal\'{e},~L.; Schulten,~K. {Scalable molecular
  dynamics with NAMD.} \emph{Journal of computational chemistry} \textbf{2005},
  \emph{26}, 1781--802\relax
\mciteBstWouldAddEndPuncttrue
\mciteSetBstMidEndSepPunct{\mcitedefaultmidpunct}
{\mcitedefaultendpunct}{\mcitedefaultseppunct}\relax
\EndOfBibitem
\bibitem[MacKerell \latin{et~al.}(1998)MacKerell, Bashford, Bellott, Dunbrack,
  Evanseck, Field, Fischer, Gao, Guo, Ha, Joseph-McCarthy, Kuchnir, Kuczera,
  Lau, Mattos, Michnick, Ngo, Nguyen, Prodhom, Reiher, Roux, Schlenkrich,
  Smith, Stote, Straub, Watanabe, Wi\'{o}rkiewicz-Kuczera, Yin, and
  Karplus]{MacKerell1998}
MacKerell,~A.~D. \latin{et~al.}  {All-atom empirical potential for molecular
  modeling and dynamics studies of proteins.} \emph{The journal of physical
  chemistry. B} \textbf{1998}, \emph{102}, 3586--616\relax
\mciteBstWouldAddEndPuncttrue
\mciteSetBstMidEndSepPunct{\mcitedefaultmidpunct}
{\mcitedefaultendpunct}{\mcitedefaultseppunct}\relax
\EndOfBibitem
\bibitem[MacKerell \latin{et~al.}(2004)MacKerell, Feig, and
  Brooks]{MacKerell2004a}
MacKerell,~A.~D.; Feig,~M.; Brooks,~C.~L. {Improved treatment of the protein
  backbone in empirical force fields}. \emph{J. Am. Chem. Soc.} \textbf{2004},
  \emph{126}, 698--699\relax
\mciteBstWouldAddEndPuncttrue
\mciteSetBstMidEndSepPunct{\mcitedefaultmidpunct}
{\mcitedefaultendpunct}{\mcitedefaultseppunct}\relax
\EndOfBibitem
\bibitem[Klauda \latin{et~al.}(2010)Klauda, Venable, Freites, O'Connor, Tobias,
  Mondragon-Ramirez, Vorobyov, MacKerell, and Pastor]{Klauda2010}
Klauda,~J.~B.; Venable,~R.~M.; Freites,~J.~A.; O'Connor,~J.~W.; Tobias,~D.~J.;
  Mondragon-Ramirez,~C.; Vorobyov,~I.; MacKerell,~A.~D.; Pastor,~R.~W. {Update
  of the CHARMM all-atom additive force field for lipids: validation on six
  lipid types.} \emph{J. Phys. Chem. B} \textbf{2010}, \emph{114},
  7830--7843\relax
\mciteBstWouldAddEndPuncttrue
\mciteSetBstMidEndSepPunct{\mcitedefaultmidpunct}
{\mcitedefaultendpunct}{\mcitedefaultseppunct}\relax
\EndOfBibitem
\bibitem[Lim \latin{et~al.}(2012)Lim, Rogaski, and Klauda]{Lim2012}
Lim,~J.~B.; Rogaski,~B.; Klauda,~J.~B. {Update of the cholesterol force field
  parameters in CHARMM}. \emph{J. Phys. Chem. B} \textbf{2012}, \emph{116},
  203--210\relax
\mciteBstWouldAddEndPuncttrue
\mciteSetBstMidEndSepPunct{\mcitedefaultmidpunct}
{\mcitedefaultendpunct}{\mcitedefaultseppunct}\relax
\EndOfBibitem
\bibitem[Essman \latin{et~al.}(1995)Essman, Perela, Berkowitz, Darden, and
  Pedersen]{Essman1995}
Essman,~U.; Perela,~L.; Berkowitz,~M.~L.; Darden,~T.; Pedersen,~L.~G. {A smooth
  particle mesh \{E\}wald method}. \emph{J. Chem. Phys.} \textbf{1995},
  \emph{103}, 8577--8592\relax
\mciteBstWouldAddEndPuncttrue
\mciteSetBstMidEndSepPunct{\mcitedefaultmidpunct}
{\mcitedefaultendpunct}{\mcitedefaultseppunct}\relax
\EndOfBibitem
\bibitem[Marsh(2010)]{Marsh2010}
Marsh,~D. {Liquid-ordered phases induced by cholesterol: A compendium of binary
  phase diagrams}. \emph{Biochimica et Biophysica Acta - Biomembranes}
  \textbf{2010}, \emph{1798}, 688--699\relax
\mciteBstWouldAddEndPuncttrue
\mciteSetBstMidEndSepPunct{\mcitedefaultmidpunct}
{\mcitedefaultendpunct}{\mcitedefaultseppunct}\relax
\EndOfBibitem
\bibitem[Gater \latin{et~al.}(2014)Gater, Saurel, Iordanov, Liu, Cherezov, and
  Milon]{Gater2014}
Gater,~D.~L.; Saurel,~O.; Iordanov,~I.; Liu,~W.; Cherezov,~V.; Milon,~A. {Two
  Classes of Cholesterol Binding Sites for the $\beta$2AR Revealed by
  Thermostability and NMR}. \emph{Biophysj} \textbf{2014}, \emph{107},
  2305--2312\relax
\mciteBstWouldAddEndPuncttrue
\mciteSetBstMidEndSepPunct{\mcitedefaultmidpunct}
{\mcitedefaultendpunct}{\mcitedefaultseppunct}\relax
\EndOfBibitem
\bibitem[Chodera \latin{et~al.}(2011)Chodera, Mobley, Shirts, Dixon, Branson,
  and Pande]{Chodera2011}
Chodera,~J.~D.; Mobley,~D.~L.; Shirts,~M.~R.; Dixon,~R.~W.; Branson,~K.;
  Pande,~V.~S. {Alchemical free energy methods for drug discovery: progress and
  challenges.} \emph{Current opinion in structural biology} \textbf{2011},
  \emph{21}, 150--160\relax
\mciteBstWouldAddEndPuncttrue
\mciteSetBstMidEndSepPunct{\mcitedefaultmidpunct}
{\mcitedefaultendpunct}{\mcitedefaultseppunct}\relax
\EndOfBibitem
\bibitem[Wang \latin{et~al.}(2015)Wang, Wu, Deng, Kim, Pierce, Krilov, Lupyan,
  Robinson, Dahlgren, Greenwood, Romero, Masse, Knight, Steinbrecher, Beuming,
  Damm, Harder, Sherman, Brewer, Wester, Murcko, Frye, Farid, Lin, Mobley,
  Jorgensen, Berne, Friesner, and Abel]{Wang2015}
Wang,~L. \latin{et~al.}  Accurate and Reliable Prediction of Relative Ligand
  Binding Potency in Prospective Drug Discovery by Way of a Modern Free-Energy
  Calculation Protocol and Force Field. \emph{J. Am. Chem. Soc.} \textbf{2015},
  \emph{137}, 2695--2703\relax
\mciteBstWouldAddEndPuncttrue
\mciteSetBstMidEndSepPunct{\mcitedefaultmidpunct}
{\mcitedefaultendpunct}{\mcitedefaultseppunct}\relax
\EndOfBibitem
\bibitem[Mokrab \latin{et~al.}(2009)Mokrab, Stevens, and Mizuguchi]{Mokrab2009}
Mokrab,~Y.; Stevens,~T.~J.; Mizuguchi,~K. Lipophobicity and the residue
  environments of the transmembrane alpha-helical bundle. \emph{Proteins}
  \textbf{2009}, \emph{74}, 32--49\relax
\mciteBstWouldAddEndPuncttrue
\mciteSetBstMidEndSepPunct{\mcitedefaultmidpunct}
{\mcitedefaultendpunct}{\mcitedefaultseppunct}\relax
\EndOfBibitem
\bibitem[Duneau and Sturgis(2013)Duneau, and Sturgis]{Duneau2013}
Duneau,~J.-P.; Sturgis,~J.~N. Lateral organization of biological membranes:
  role of long-range interactions. \emph{European biophysics journal : EBJ}
  \textbf{2013}, \emph{42}, 843--850\relax
\mciteBstWouldAddEndPuncttrue
\mciteSetBstMidEndSepPunct{\mcitedefaultmidpunct}
{\mcitedefaultendpunct}{\mcitedefaultseppunct}\relax
\EndOfBibitem
\bibitem[Duneau \latin{et~al.}(2017)Duneau, Khao, and Sturgis]{Duneau2017}
Duneau,~J.-P.; Khao,~J.; Sturgis,~J.~N. Lipid perturbation by membrane proteins
  and the lipophobic effect. \emph{Biochimica et biophysica acta}
  \textbf{2017}, \emph{1859}, 126--134\relax
\mciteBstWouldAddEndPuncttrue
\mciteSetBstMidEndSepPunct{\mcitedefaultmidpunct}
{\mcitedefaultendpunct}{\mcitedefaultseppunct}\relax
\EndOfBibitem
\end{mcitethebibliography}

\end{document}